\documentclass[twocolumn]{aastex63}
\usepackage{txfonts}
\usepackage{graphicx}
\usepackage{multirow}
\usepackage{rotating}
\usepackage{verbatim}
\usepackage{url}
\usepackage{dcolumn}
\usepackage[utf8]{inputenc}

\newcommand{\Nstars}{427 }
\newcommand{\Nstarsfield}{321 }
\newcommand{\NstarsScoCen}{24 }
\newcommand{\NstarspreLASSO}{82}
\newcommand{\NstarsKP}{346 }
\newcommand{\NstarsUH}{98 }
\newcommand{\Nrepeats}{17 }
\newcommand{\NobsVIC}{432 }
\newcommand{\Nhostcandidates}{111 } 
\newcommand{\NhostsKPUH}{four } 
\newcommand{\Ntriples}{10} 
\newcommand{\Ncomps}{121 } 
\newcommand{\NcompsKP}{100 } 
\newcommand{\NcompsUH}{25 } 
\newcommand{\NhostsVIC}{107 } 
\newcommand{\NcompsVIC}{118 } 
\newcommand{\NcompsVICdet}{100 } 
\newcommand{\NcompsNOVICdet}{17 } 
\newcommand{\NcompsGaia}{75 } 
\newcommand{\NcompsGaiaunknown}{59 }
\newcommand{\Ncompsprlxs}{62 } 
\newcommand{\NcompsGaiabound}{50 } 
\newcommand{\NcompsLitbound}{nine } 
\newcommand{\Ncompsbound}{62 } 
\newcommand{\Ncompsbackground}{14 } 
\newcommand{\Ncompsunknown}{45 } 
\newcommand{\NtriplesNOTbkg}{four} 
\newcommand{\Ncompscat}{38 } 

\newcommand\note[1]{\textbf{\color{red}#1}}   
\newcommand\todo[1]{\textbf{\noindent \color{red} $\square$ #1}}
\newcommand\new[1]{{\color{blue}#1}}          
\newcommand\neww[1]{{\color{blue}#1}}          
\renewcommand\note[1]{\hspace{-1sp}}             
\renewcommand\todo[1]{\hspace{-1sp}}             
\renewcommand\new[1]{\color{black}#1}   
\renewcommand\neww[1]{\color{black}#1}   

\graphicspath{{./}{figures/}}
\usepackage{float}
\usepackage[caption = false]{subfig}
\usepackage{hyperref}

\shorttitle{LASSO Robo-AO Survey}
\shortauthors{Salama et al.}

\begin{document}

\title{Large Adaptive Optics Survey for Substellar Objects (LASSO) Around Young, Nearby, Low-mass Stars with Robo-AO}


\correspondingauthor{Ma\"issa Salama}
\email{msalama@hawaii.edu}

\author[0000-0002-5082-6332]{Ma\"issa Salama}
\affiliation{Institute for Astronomy, University of Hawai`i at M\={a}noa, Hilo, HI 96720, USA}

\author[0000-0002-8439-7767]{James Ou}
\affiliation{Institute for Astronomy, University of Hawai`i at M\={a}noa, Hilo, HI 96720, USA}

\author[0000-0002-1917-9157]{Christoph Baranec}
\affiliation{Institute for Astronomy, University of Hawai`i at M\={a}noa, Hilo, HI 96720, USA}

\author[0000-0003-2232-7664]{Michael C. Liu}
\affiliation{Institute for Astronomy, University of Hawai`i at M\={a}noa, Honolulu, HI 96822, USA}

\author[0000-0003-2649-2288]{Brendan P. Bowler}
\affiliation{Department of Astronomy, The University of Texas at Austin, Austin, TX 78712, USA}

\author{Paul Barnes}
\affiliation{Institute for Astronomy, University of Hawai`i at M\={a}noa, Hilo, HI 96720, USA}

\author{Morgan Bonnet}
\affiliation{Institute for Astronomy, University of Hawai`i at M\={a}noa, Honolulu, HI 96822, USA}

\author[0000-0002-8462-0703]{Mark Chun}
\affiliation{Institute for Astronomy, University of Hawai`i at M\={a}noa, Hilo, HI 96720, USA}

\author[0000-0001-5060-8733]{Dmitry A. Duev}
\affiliation{Division of Physics, Mathematics, and Astronomy, California Institute of Technology, Pasadena, CA 91125, USA}

\author{Sean Goebel}
\affiliation{Institute for Astronomy, University of Hawai`i at M\={a}noa, Hilo, HI 96720, USA}

\author{Don Hall}
\altaffiliation{Author is deceased.}
\affiliation{Institute for Astronomy, University of Hawai`i at M\={a}noa, Hilo, HI 96720, USA}

\author{Shane Jacobson}
\affiliation{Institute for Astronomy, University of Hawai`i at M\={a}noa, Hilo, HI 96720, USA}

\author[0000-0003-0054-2953]{Rebecca Jensen-Clem}
\affiliation{Astronomy \& Astrophysics Department, University of California, Santa Cruz, CA 95064, USA}

\author[0000-0001-9380-6457]{Nicholas M. Law}
\affiliation{Department of Physics and Astronomy, University of North Carolina at Chapel Hill, Chapel Hill, NC 27599-3255, USA}

\author{Charles Lockhart}
\affiliation{Institute for Astronomy, University of Hawai`i at M\={a}noa, Hilo, HI 96720, USA}

\author[0000-0002-0387-370X]{Reed Riddle}
\affiliation{Division of Physics, Mathematics, and Astronomy, California Institute of Technology, Pasadena, CA 91125, USA}

\author{Heather Situ}
\affiliation{Institute for Astronomy, University of Hawai`i at M\={a}noa, Honolulu, HI 96822, USA}

\author{Eric Warmbier}
\affiliation{Institute for Astronomy, University of Hawai`i at M\={a}noa, Hilo, HI 96720, USA}

\author[0000-0002-3726-4881]{Zhoujian Zhang}
\affiliation{Institute for Astronomy, University of Hawai`i at M\={a}noa, Honolulu, HI 96822, USA}


\begin{abstract}
\new{We present results from} the Large Adaptive optics Survey for Substellar Objects (LASSO)\new{, where the goal} is to directly image new substellar companions ($<$70 M$_{Jup}$) at wide orbital separations ($\gtrsim$50~AU) around young ($\lesssim$300 Myrs), nearby ($<$100 pc), low-mass (\new{$\approx$}0.1--0.8 M$_{\odot}$) stars. We report on \Nstars young stars imaged in the visible (\textit{i'}) and near-infrared (\textit{J} or \textit{H}) simultaneously with Robo-AO on the Kitt Peak 2.1-m telescope and later the Maunakea U\new{niversity of} H\new{awaii} 2.2-m telescope. To undertake the observations, we commissioned a new infrared camera \new{for} Robo-AO that uses a low-noise high-speed SAPHIRA \new{avalanche photodiode} detector. We detected \Ncomps companion candidates around \Nhostcandidates stars, of which \Ncompsbound companions are physically associated based on \textit{Gaia} DR2 parallaxes and proper motions, another \Ncompsunknown require follow-up observations to confirm physical association, and \Ncompsbackground are background objects. The companion separations range from 2--1101 AU and reach contrast ratios of 7.7 magnitudes in the near infrared compared to the primary. The majority of confirmed and pending candidates are stellar companions, with $\sim$5 being potentially substellar and requiring follow-up observations for confirmation. We also detected a 43$\pm$9 M$_{Jup}$ and an 81$\pm$5 M$_{Jup}$ companion that were previously reported. We found 34 of our targets have acceleration measurements \new{detected using} \textit{Hipparcos-Gaia} proper motions. Of \new{those, 58$^{+12}_{-14}$\% of the} 12 stars \new{with imaged} companion candidate\new{s have} significant accelerations ($\chi^2 >11.8$), while only \new{23$^{+11}_{-6}$}\% of the \new{remaining} 22 stars with no detected companion have significant accelerations. The significance of the acceleration decreases with increasing companion separation. These young accelerating low-mass stars with companions will eventually yield dynamical masses with future orbit monitoring.
\end{abstract}

\keywords{stars: low-mass - binaries: visual - instrumentation: adaptive optics - techniques: high angular resolution - methods: observational - surveys - brown dwarfs}

\section{Introduction} \label{sec:intro}

Over the past three decades, our knowledge of planetary systems has expanded from just our Solar System to a multitude of planetary architectures. The Kepler mission detected thousands of exoplanets in close-in orbits ($\lesssim$1~AU) as they transit their host star. The radial velocity method has been used to discover thousands of exoplanets out to slightly farther orbits ($\lesssim$5~AU). Direct imaging helped identify a complementary population of substellar companions (2--75~M$_{Jup}$), namely planets and brown dwarfs, at large projected separations ($\sim$5--8000~AU). The existence of these companions at wide separations has played a critical role in shaping theories about the formation and migration of brown dwarfs and planets, through the development of mechanisms such as core/pebble accretion \citep{LambJoh12}, disk instability \citep{Durisen07,KrattLod16}, cloud fragmentation \citep{Bate03}, and dynamical scattering \citep{Veras09} on various timescales and orbital separations. These mechanisms predict correlations between the presence of a wide-orbit companion and certain environmental characteristics, such as the presence or absence of other companions, circumstellar disk morphologies, and the eccentricity of the companion's orbit, which can be compared to observational studies. The IAU currently defines the boundary between brown dwarfs and planets at 13~M$_{Jup}$, but it remains unresolved whether this is an artificial boundary or reflective of a natural division that links observational properties of these objects to dominant formation mechanisms \citep{Chabrier14,Schlaufman18}. 

Population demographic studies are necessary to search for trends in the orbital architectures, primary and companion masses, ages, and environments of systems with wide-orbit substellar companions and to clarify the boundary between brown dwarfs and massive exoplanets. However, the number of discoveries so far has limited the statistical analysis of these trends. Large exoplanet imaging searches, each on the order of hundreds of targets, have discovered between 0 and 4 substellar companions, bringing the total detections to $<$20~objects in the planetary-mass regime ($\lesssim$13~M$_{Jup}$) and another $\sim$100 in the brown dwarf regime \citep{Deacon14,Bowler16,BowlerNielson18,Baron19}. \cite{Bowler20} conducted one of the first population demographic studies comparing the eccentricity distributions of substellar companion orbits over various parameters (e.g. mass, separation, age). With a sample size of 27 substellar companions with orbital measurements, they found differences in the peaks of the distributions but could not constrain the exact shape of the distributions. Population studies with larger sample sizes are needed to better understand how these substellar objects form and evolve, as well as determine the natural boundary distinguishing brown dwarfs from massive exoplanets. A key step is to conduct a survey large enough to greatly boost the detections of these rare objects. This will allow us to perform more detailed population studies to test formation and evolution models.

\new{Many of the early direct imaging surveys focused on massive stars because of the better AO corrected image quality and easier identification of young stars. However, low-mass stars are by far the most abundant stars in the galaxy, comprising roughly 75\% of all stars \citep{Bochanski10}. Several direct imaging surveys focusing on low-mass stars have now been conducted. For example, the Planets Around Low-Mass Stars (PALMS; \citealt{Bowler15}) survey observed 122 young M dwarfs with Keck/NIRC2 and Subaru/HiCIAO and detected 4 brown dwarf companions and no planetary companions. The M-dwArf Statistical Survey for direct Imaging of massiVe Exoplanets (MASSIVE; \citealt{Lannier16}) observed 58 young and nearby M dwarfs and did not detect new substellar companions. The Planet Search around Young-associations M dwarfs (PSYM-WIDE; \citealt{Naud17}), a deep seeing-limited survey, observed 95 stars with Gemini/GMOS and discovered one planetary companion.}

Most stars are believed to have formed as part of a multiple system from the collapse and fragmentation of cloud cores \citep{Larson02}. However, the frequency of multiple systems has been observed to decrease with age \citep{Duchene13}, implying that as a stellar system evolves dynamical interactions cause the ejection of companions \citep{Reipurth14}. Thus, multiplicity statistics of young stars are useful for placing boundary conditions in evolutionary models exploring companion loss processes. 

Adaptive optics (AO) technology, which corrects for the blurring effect of the atmosphere, has enabled the discovery of many wide-orbit substellar companions by direct imaging. We are conducting a companion survey using Robo-AO, a robotic laser adaptive optics instrument \citep{Baranec14} at the Kitt Peak 2.1-m and Maunakea UH 2.2-m telescopes. Robo-AO's infrared science camera is equipped with a SA\new{PH}IRA (Selex Avalanche Photodiode for High-speed Infrared Array) detector \citep{Baranec15}. This is a new type of infrared detector using electron-avalanche mechanisms to boost the signal while keeping the read noise fixed. As such, our survey is also testing the sensitivity and on-sky performance of these detectors. 

We report here on the results from our observations as part of the Large Adaptive optics Survey for Substellar Objects (LASSO). The goal of LASSO is to search for wide-orbit (50--1500 AU), substellar companions around young ($\lesssim$300 Myrs), nearby ($<$100 pc), low-mass (0.1--0.8 M$_{\odot}$) stars. In Section \ref{sec:LASSO} we introduce the LASSO survey, target selection (\S\ref{sec:LASSO:target_list}), Robo-AO instrument (\S\ref{sec:LASSO:RoboAO}), observations (\S\ref{sec:LASSO:obs}), data reduction (\S\ref{sec:LASSO:reduction}), and companion detection method (\S\ref{sec:LASSO:comp_detect}). In Section \ref{sec:results} we report the results of our observations, physical association determination (\S\ref{sec:bound}) and optical-infrared colors (\S\ref{sec:colors}). In Section \ref{sec:Discussion} we discuss and analyze triple systems (\S\ref{sec:triples}), accelerating stars (\S\ref{sec:accelerations}), substellar objects (\S\ref{sec:substellar}), and survey yields \& expectations (\S\ref{sec:yields}), and in Section \ref{sec:Conclusion} we summarize our main conclusions.

\section{Survey and Observations} \label{sec:LASSO}

The objective of LASSO is to find new substellar companions at wide separations in order to carry out population studies of these rare objects. We surveyed young, nearby, low-mass stars because substellar objects are brighter when they are younger; it is easier to resolve objects at small physical separations when they are closer to us, and we are more sensitive to lower-mass companions in systems with lower mass primaries. \new{This is an ongoing survey with more observations to come from a prioritized sample.}

\subsection{LASSO Target Selection} \label{sec:LASSO:target_list}

We used the Cool Dwarf Catalog (CDC; \citealt{Muirhead18}) as our starting sample. Its purpose is to identify cool dwarf targets for the Transiting Exoplanet Survey Satellite (TESS). Because young low-mass stars have chromospheric activity due to strong magnetic dynamos caused by their deep convective envelopes and differential rotation, they can be identified by excesses in the UV. In order to select for young stars, we cross-matched the CDC with the Galaxy Evolution Explorer (GALEX; \citealt{Martin05,Morrissey07}). Following \cite{Rodriguez13}, we applied the following selection criteria:
\begin{equation}
NUV - W1 \leq 12.5\ mag
\label{eq:col1}
\end{equation}
\begin{equation}
 J - W2 \geq 0.8\ mag
\label{eq:col2}
\end{equation}
\begin{equation}
NUV - W1 < 7 \times (J - W2) + 5.5\ mag
\label{eq:col3}
\end{equation}
Where NUV photometry is from GALEX, \textit{J} magnitudes are from the Two Micron All Sky Survey (2MASS; \citealt{Skrutskie06}), and W1 (3.4~$\mu$m) and W2 (4.6~$\mu$m) are from the Wide-field Infrared Survey Explorer (WISE; \citealt{Wright10}). The remaining targets were then cross-matched with \textit{Gaia} DR2 \citep{GaiaCollab16,GaiaCollab18b} to filter by measured proper motion and distance. Only targets within 100~pc were selected. Finally, we selected only targets observable from Kitt Peak (KP), and later Maunakea (MK), by limiting to targets with declination $> -30^{\circ}$ and $-35^{\circ}$, respectively. Targets with $i' \leq 15$ mag were selected, or $V \leq 17$ mag when no \textit{i'}-band measurement was available in the CDC. 

\begin{figure}[ht!]
\centering
\includegraphics[width=240pt]{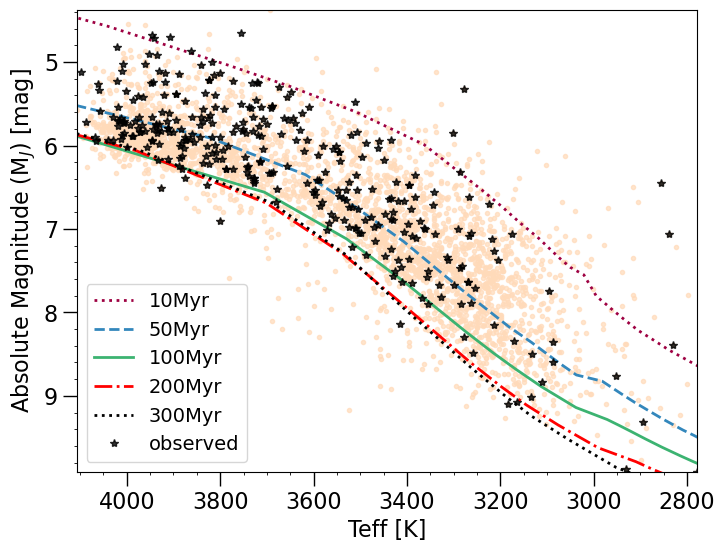}
\caption{LASSO young field target sample \new{(light orange dots)} with isochrones from the \cite{Baraffe15} evolutionary models overlayed. The effective temperatures and $\textit{J}$-band magnitudes are from the CDC, and \textit{Gaia} DR2 was used to determine $\textit{J}$-band absolute  magnitudes. The estimated age range of our target sample is $\sim$10--\new{300}~Myr. \new{Observed stars reported in this paper are shown as black stars.}
    \label{fig:targets}}
\end{figure}

The resulting target list comprises 2,787 stars for Kitt Peak and 3,291 stars for Maunakea. Properties of our sample are shown in Figure~\ref{fig:targets}, with evolutionary isochrones overlayed. Most targets are estimated to be between $\sim$10--\new{300}~Myr old\new{, though no detailed independent age estimates were performed and some targets may be older than expected (see discussion in \S \ref{sec:yields})}. The targets are predominantly M dwarfs with some late K-type stars. They span temperatures of $\sim$3000--4000~K, and masses of $\sim$0.1--0.8~$M_{\odot}$. We report Robo-AO imaging of \new{\Nstarsfield} stars from this sample in this paper. \new{The observed stars were selected by the Robo-AO queue system (\S \ref{sec:LASSO:RoboAO}).}

\subsubsection{Sco-Cen Targets}\label{sec:LASSO:ScoCensample}
In order to compare a younger sample of targets to our young field sample, after moving Robo-AO to Maunakea in 2019 we added targets from the Scorpius-Centaurus (Sco-Cen) association, which is the youngest and nearest OB association observable from the Northern Hemisphere \citep{deZeeuw99,WrightMamajek18}. The Sco-Cen association is farther than our young field sample, at an average distance of 140~pc, but it is younger, with an estimated age range of 5--15~Myr. We will thus be more sensitive to lower mass objects than our field sample, but at slightly larger orbital separations. We used the \cite{VillaVelez18} sample that isolated the pre-main sequence population of stars in the Sco-Cen association using \textit{Gaia} DR2 and applied the same color selection criteria as used by the CDC to select low-mass cool stars:
\begin{equation}
 V - J > 2.7\ mag
\end{equation}
\begin{equation}
M_V > 2.2 \times (V - J) - 2.0\ mag
\end{equation}
We also applied the same declination and magnitude cuts (Dec $> -35^{\circ}$ and $i' \leq 15$ mag) as for our field sample. This additional sample consists of 668 targets, $\sim$60\% of which are in the Upper Scorpius region, the youngest (5--15~Myr) subgroup of the association. We report Robo-AO imaging for \NstarsScoCen stars from this sample in this paper.

\subsubsection{Pre-LASSO targets} \label{sec:LASSO:preLASSO}
Before we finalized the target list above, and prior to the CDC and \textit{Gaia} data releases, we used a preliminary target list of young active M-dwarfs. These targets were selected from the cross-match of color selected samples from \cite{Frith13} and \cite{Haakonsen09}, described in \cite{Bowler19}. \new{In this paper we also report on Robo-AO imaging of 82 stars selected from this preliminary target list which} would not have been part of our LASSO list for the following reasons: \new{19} were not part of the CDC, \new{42} did not yield a match in GALEX, and \new{21} did not satisfy the selection criteria described in Equations \ref{eq:col1}--\ref{eq:col3}. \new{31 additional stars were also selected and observed from this preliminary target list, but would} have been part of our LASSO sample \new{and thus were included in our LASSO count discussed in \S\ref{sec:LASSO:target_list}}.

\new{Table \ref{tab:obs_lists_summary} summarizes the number of stars reported in this paper and which list they come from.}

\begin{table}[h]
\centering
\caption{\new{Observed Stars Target Lists}
    \label{tab:obs_lists_summary}}
\begin{tabular}{lc}
& \new{\textbf{Number of}} \\
\new{\textbf{Target List}} & \new{\textbf{stars observed}} \\
\hline
\new{LASSO young field late-K/M dwarfs} & \new{321} \\
\new{Pre-LASSO young field M dwarfs} & \new{82} \\
\new{Sco-Cen association late-K/M dwarfs} & \new{24} \\
\hline
\new{Total} & \new{427} \\
\hline
\end{tabular}
\end{table}

\subsection{Robo-AO Instrument} \label{sec:LASSO:RoboAO}
We conducted the observations with the Robo-AO instrument first at the Kitt Peak National Observatory 2.1-m telescope in Arizona, and later at the UH 2.2-m telescope on Maunakea, Hawai`i \citep{Baranec14,Salama16,JensenClem18,Salama18}. Robo-AO is equipped with both visible and infrared science cameras with a dichroic mirror simultaneously sending wavelengths shorter than $\lambda$=950 nm to the visible camera and the longer wavelengths to the infrared camera. Robo-AO uses a Rayleigh-scattering laser guide star with a line of sight focus at $\sim$10~km. The AO system runs at a rate of 1.2~kHz to correct high-order wavefront aberrations. In order to correct the tip-tilt motion of the star (not sensed by the laser guide star), we process the images with post-facto shift-and-add, with data taken at 20~Hz in the infrared and 8.6~Hz in the visible. See \cite{JensenClem18} for more details. 

We installed Robo-AO on the 2.1-m telescope on Kitt Peak, Arizona, from November 2015 to June 2018. We conducted our observations after commissioning the infrared camera in November 2016 and while we were still testing and characterizing the performance of the detector and integrating the readout software within our automated observing routines. Each observation consists of a 5-minute exposure with 1-2 minutes of overhead due to telescope slewing and pointing, and laser guide star acquisition. The Robo-AO intelligent observing queue is described in \cite{Riddle14}.

\subsubsection{Infrared Camera} \label{sec:LASSO:RoboAO:SAPHIRA}

In order to extend the scope of observable objects to much cooler and lower-mass objects (brown dwarfs and massive exoplanets) we added an infrared science camera with a SAPHIRA detector \citep{Salama18}. Selex Avalanche Photodiode for HgCdTe InfraRed Array (SAPHIRA) detectors \citep{Finger14} provide photon-counting technology at infrared wavelengths \new{\citep{Atkinson18}}. SAPHIRA detectors make use of electron-avalanche mechanisms within each pixel to effectively multiply the signal without increasing the read noise and thus improving the Signal-to-Noise Ratio (SNR). SAPHIRA detectors allow for almost noiseless signal amplification and ultra-low dark currents \citep{Atkinson17}, which is especially beneficial for ground-based astronomical observations of photon-starved targets (e.g., \new{\citealt{Goebel18}}; \citealt{Hippler20,Bond20}) . This type of high-speed detector is particularly useful to minimize the degrading effect of tip-tilt displacement on image quality by taking multiple short-exposure images while adding negligible noise \citep{JensenClem18}. 

The infrared camera filter is located inside of the camera dewar and maintained at a temperature of 85K. After testing different filters in the lab, calculating sensitivity limits, and changing the filter on-sky at Kitt Peak, we determined that we were most sensitive to substellar objects in the \textit{J} band due to the high thermal background at longer wavelengths. Substellar objects are brighter in the \textit{H} band, and because thermal background is lower due to the colder temperatures on Maunakea, we installed an \textit{H}-band filter when we moved to the UH~2.2-m telescope.

\subsubsection{Visible Camera} \label{sec:LASSO:RoboAO:VIC}

Robo-AO also has an EMCCD visible-light science camera with a filter wheel. The filter wheel includes the \textit{g', r', i'}, and \textit{z'} filters as well as a long-pass ``lp600" filter, allowing wavelengths longer than 600~nm through. We carried out our observations in the \textit{i'}-band for optimal image sharpness. The characteristics of both science cameras are summarized in Table \ref{tb:cam_specs}.

\begin{table}[h]
\centering
\caption{Robo-AO science cameras
    \label{tb:cam_specs}}
\begin{tabular}{lcc}
& \textbf{Visible Camera} & \textbf{Infrared Camera} \\
\hline
Detector & EMCCD & SAPHIRA \\
Wavelengths & 400 -- 950 nm &  0.8 -- 2.5 $\mu$m \\
Format & \new{1024 $\times$ 1024 pixels} & \new{320 $\times$ 256 pixels} \\
Pixel size & 13 $\mu$m & 24 $\mu$m \\
Field of view & 36$\arcsec \times$ 36$\arcsec$ (KP) & 20.5$\arcsec \times$ 16.5$\arcsec$ (KP) \\
 & 26$\arcsec \times$ 26$\arcsec$ (MK) & 14.5$\arcsec \times$ 11.5$\arcsec$ (MK) \\
Plate scale & 35 mas/pixel (KP) & 64 mas/pixel (KP) \\
 & 25 mas/pixel (MK) & 46 mas/pixel (MK) \\
Filters & \textit{i'} & \textit{J} (KP), \textit{H} (MK)\\
Sampling rate & 8.6~Hz & 20~Hz \\
\hline
\end{tabular}
\end{table}

\subsection{Observations} \label{sec:LASSO:obs}

We observed a total of \Nstars stars with Robo-AO. \NstarsKP observations were carried out on the 2.1-m telescope on Kitt Peak, Arizona from 2017 June to 2018 June. \NstarsUH observations were conducted in 2019 May and October on the UH~2.2-m telescope on Maunakea, \Nrepeats of which we had also previously observed at Kitt Peak. We obtained simultaneous images in the \textit{i'} band and \textit{J} band (on Kitt Peak) or \textit{i'} band and \textit{H} band (on Maunakea) for \NobsVIC of our observations. \new{We did not capture simultaneous visible images for twelve stars, as the infrared camera software was not yet fully integrated with the Robo-AO system.} The median measured seeing in the \textit{i'}-band was $1.53\arcsec \pm 0.26\arcsec$ and $0.97\arcsec \pm 0.24\arcsec$ at Kitt Peak and Maunakea, respectively. Table \ref{tab:obs_all} in the Appendix lists all of the observed targets along with the observing conditions and achieved contrasts.

\subsection{Data Reduction} \label{sec:LASSO:reduction}

\subsubsection{Infrared camera}
The SAPHIRA camera produces raw data cubes containing the sequences of frames resulting from non-destructive readouts. We first subtract each frame from the subsequent frame to produce a differential frame, which is inherently bias corrected. To calibrate these differential frames we then subtract the sky background and divide by the flat field response. For the Robo-AO observations at the UH 2.2-m telescope, an IR source to produce flat fields had not yet been installed. Instead, we derived the per-pixel response from sky background images taken over the course of the observing program. Assuming linear response, the signal $s(t)$ returned by a pixel is the result of its response $r$, source flux over time $f_{source} \times t$, dark current (fixed pattern noise) over time $d \times t$, and bias $b$:
\begin{equation}
 s(t) = r\times f_{source} \times t + d \times t + b
\end{equation}
With fixed integration time, bias correction through differential frames, and using the median of each background as the source flux, the normalized response for each pixel can be estimated through linear regression. For consistency, we also used this process for the Kitt Peak observations. We identified pixels with poor response by sigma clipping reduced background images, and replaced these pixel values by Gaussian interpolation of surrounding pixels. This affected $\sim$2.3\% of pixels at Kitt Peak and $\sim$1\% of pixels at Maunakea. This improvement is likely due to readout electronic hardware modifications implemented while moving Robo-AO from Kitt Peak to Maunakea.

In the Kitt Peak observations we found linear artifacts that survived background subtraction and flat-fielding. These manifested as horizontal lines 32 pixels in length that appeared in random rows and were offset from the pixels in the rows above or below. Their locations corresponded to detector readout electronics and were not found in Maunakea observations. We suspect there may have been camera readout effects dependent on either hardware calibration or operating temperature, which improved after the hardware modifications and move to Maunakea. We mitigated these by subtracting a 10\% quantile value from each 32 pixel-long row segment corresponding to the readout electronics (higher quantiles oversubtracted where stars or companions were present). 

Tip-tilt motion is not sensed by the laser guide star system and is handled post-facto by a modified shift-and-add routine. We stack the calibrated differential frames using the centroid position of the brightest star. This is performed with sub-pixel precision by weighting flux contributions from each pixel in subsequent frames by their proportional overlap over the output pixels, similar to the Drizzle algorithm \citep{Fruchter02} but without shrinking the input pixels. We also produced images using shift-and-median instead of shift-and-add, which helped suppress noise artifacts not already removed by the calibrations. We selected the shift-and-median reduced image instead of the shift-and-add result for $\sim$60\% of our observations.

Additionally, we developed an optional adaptation of the GenSTAC technique \citep{Howard18} to handle low SNR frames, where target identification and centroid positioning can be unreliable in individual frames. A series of low SNR frames are stacked together until the SNR is comparable to surrounding higher quality frames. The resulting centroid is then assigned to the center frame of the stack. The positioning for all other low SNR frames in the stack are estimated by cubic spline interpolation of the centroid positions of the surrounding stack center frames and higher quality frames. Reduced images from this technique were selected for $\sim$10\% of our observations and denoted in our list of observations (Appendix Table \ref{tab:obs_all}).

\subsubsection{Visible camera}
For each raw image frame, we subtract the background and apply a flat-field correction. We then process them through the image registration pipeline (first, the ``bright-star" pipeline), which stacks the individual short exposure frames on the brightest star in the field, to correct for the tip-tilt motion not sensed by the laser guide star system. However, if the registration pipeline produces a FWHM $< \lambda/D$, then it is considered a failed registration, meaning it stacked on a bright pixel because the target was too faint. We then re-process these images through the ``faint-star" pipeline, which stacks all the frames, to create a reference frame, which is dark and flat corrected, then high-pass filtered and centered on the guide star. Each individual raw frame is then also dark and flat corrected, high-pass filtered, windowed, and finally registered to the reference frame. We also process the images through the high contrast pipeline, in order to maximize the sensitivity to detect faint companions. This pipeline consists of applying a high-pass filter to remove light from the stellar halo, then a synthetic PSF of the star is generated by the Karhunen-Loève Image Processing (KLIP) algorithm, which makes use of a PSF reference library of Robo-AO observations, and is subtracted from the observed PSF. A detailed description and performance analysis of the data reduction pipelines for the visible camera is available in \S 3 of \cite{JensenClem18}.

\subsection{Companion Detection} \label{sec:LASSO:comp_detect}

\subsubsection{Infrared camera}

For each reduced image, we generated a radial average and subtracted it from the original to remove most of the primary star light. We then visually inspected the radially-subtracted images to flag companion candidates. Next, we calculated the centroid location of the star and any companion candidates. We then measured the SNR of the companion candidate by calculating the flux of the companion candidate in a circular aperture, subtracting from it the median background flux in an annulus around the star at the same separation as the companion candidate (while masking the companion candidate itself), and dividing by the noise in that same annulus. We did this over a range of aperture radii and calculated the companion candidate flux ratio with the primary star using the aperture size corresponding to the highest SNR. In combination with visual vetting of detections in our images, we applied an SNR~$> 5$ threshold (in at least either the visible or infrared image) to report a detection as a companion candidate. The errors were calculated from the standard deviation of the measurements on the individual (pre-stacked) observations and combined with the 5~mas positional systematic uncertainty after correcting for distortion, as reported in \cite{JensenClem18}. The average measurement uncertainties are: 0.03$\arcsec$ in separation, 1.5$^{\circ}$ in PA, 0.08 mag in $\textit{i'}$-band contrast, 0.12 mag in $\textit{J}$-band contrast, and 0.20 mag in $\textit{H}$-band contrast.

\begin{figure}[ht!]
\centering
\includegraphics[width=240pt]{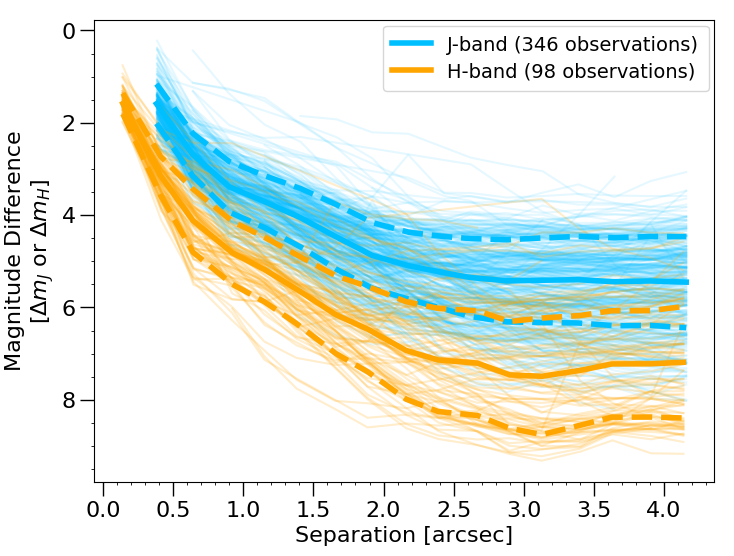}
\caption{Contrast curves from LASSO Robo-AO observations in the \textit{J}-band on the 2.1-m telescope on Kitt Peak, Arizona (blue) and in the \textit{H}-band on the 2.2-m UH telescope on Maunakea, Hawai`i (orange). The solid line is the median contrast curve and the dashed lines are $\pm$ 1 standard deviation.
    \label{fig:contrast_curve}}
\end{figure}

The sensitivity of our observations was determined through injection and recovery of fake companions in the images. For each observation, we scaled a copy of the target star to a given contrast, then injected that scaled PSF at a given separation and position angle in the image. We then subtracted the radial average of the injected image. We did this for a range of separations, position angles, and contrasts and then determined the contrast at each position where the injected companion could no longer be recovered, using an SNR threshold of 5. At each separation, we adopted the median achieved contrast over the range of position angles to generate an individual contrast curve for each target (Figure \ref{fig:contrast_curve}). For stars with companion candidates, we masked the companion when generating the contrast curve. 

\begin{figure}[ht!]
\centering
\subfloat{\includegraphics[width=240pt]{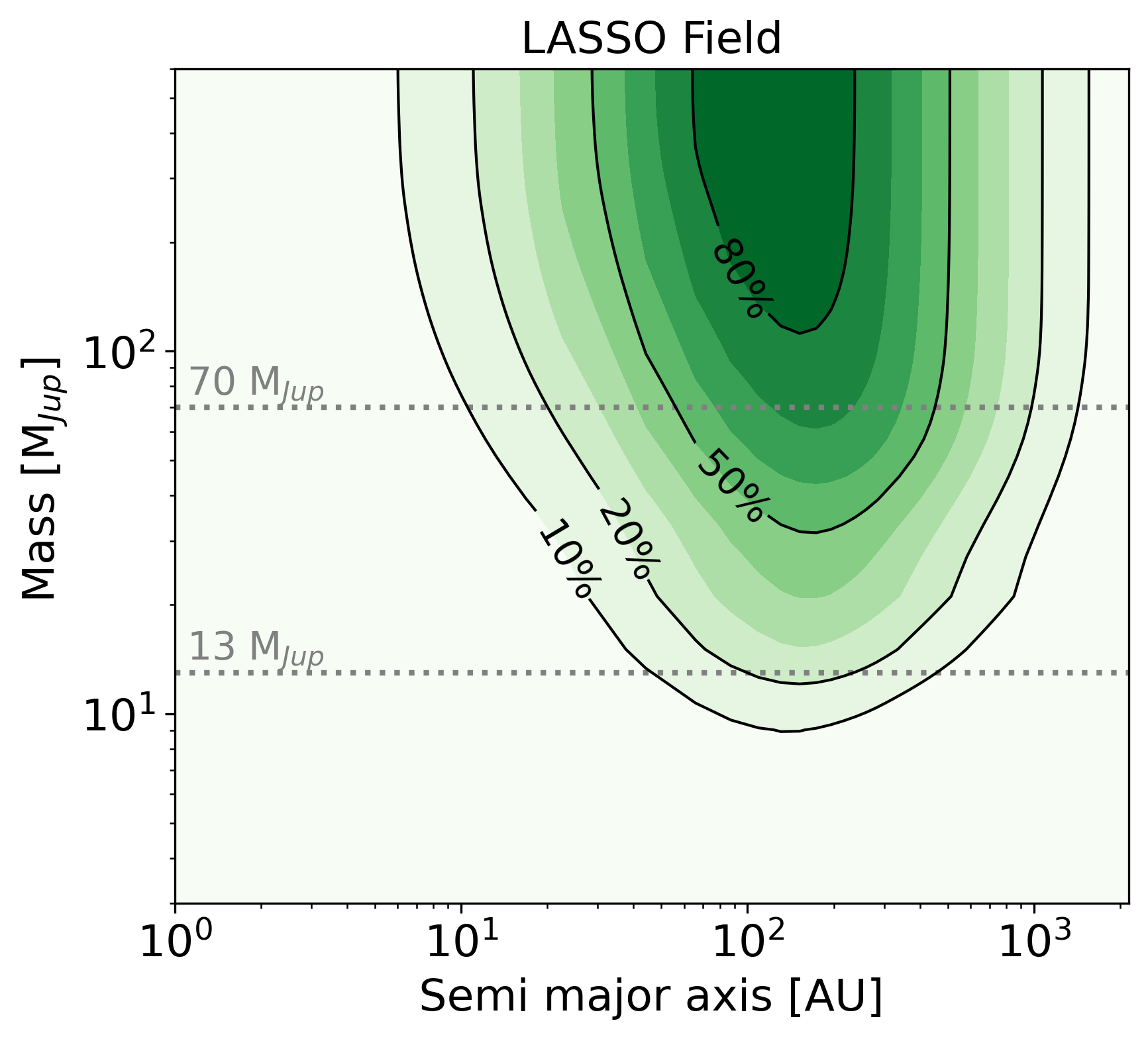}}\\
\subfloat{\includegraphics[width=240pt]{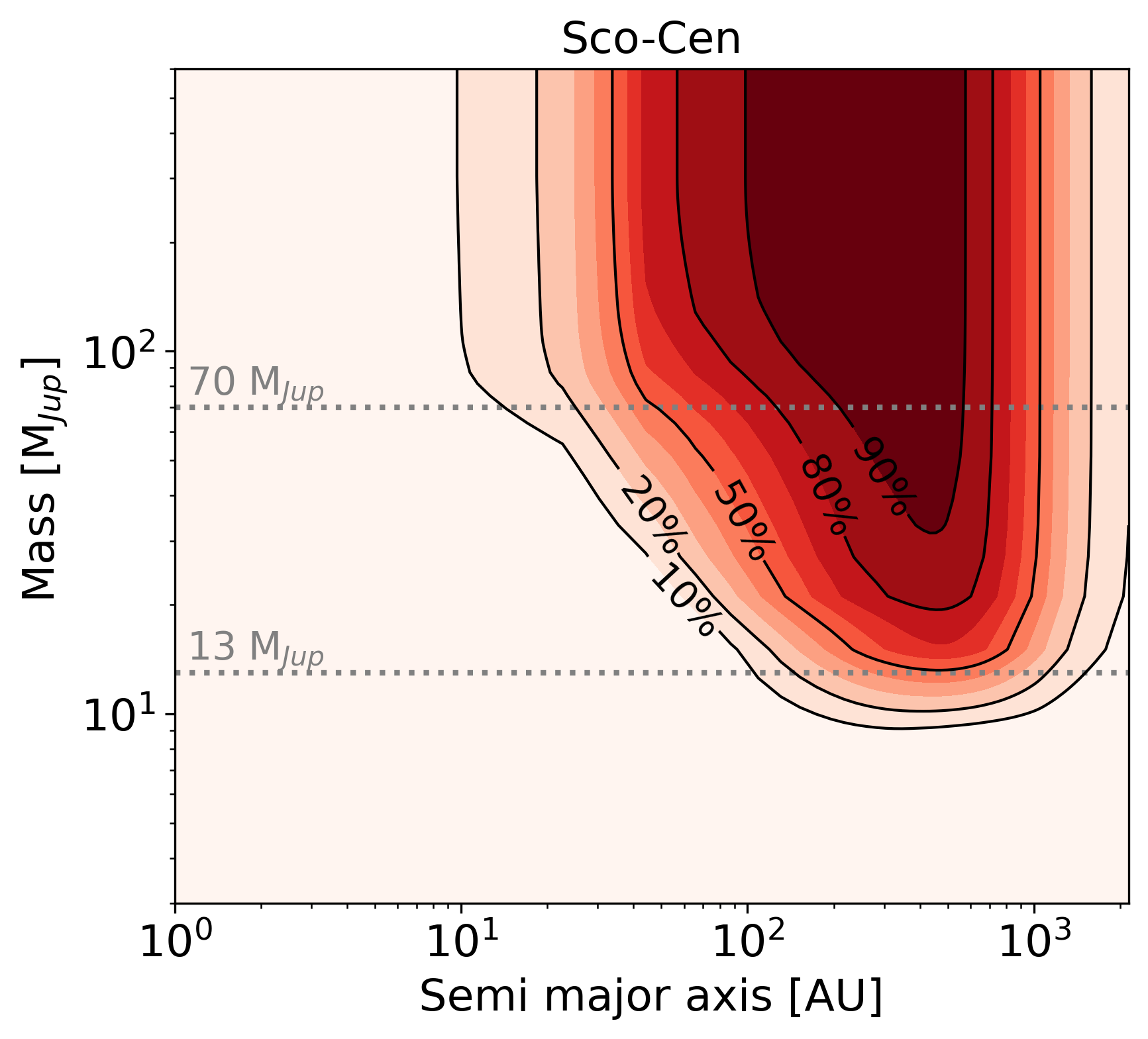}}
\caption{\new{Detection sensitivity for young field targets (\textit{top}) and Sco-Cen targets (\textit{bottom}). The contrast curves for each star was converted to physical separation in AU using the \textit{Gaia} DR2 distance, and the contrast was converted to companion mass using isochrone models over a range of ages (10--300 Myrs for field stars and 5--15 Myrs for Sco-Cen stars). The completeness was then determined using the \texttt{ExoDMC} synthetic population simulation code. } 
    \label{fig:detection_sensitivity}}
\end{figure}

\new{In order to convert our contrast curves to detection sensitivity of companion mass and physical separation, we need to take into account the distance to each primary star, its magnitude, and evolutionary models to compute companion masses. Using each observed target's \textit{Gaia} DR2 parallax, we converted the contrast curve of magnitude differences to companion absolute magnitudes as a function of projected physical separation in AU. From the absolute magnitudes and using evolutionary models from \cite{Chabrier00} and \cite{Baraffe15}, we estimated a range of 100 companion masses corresponding to a range of 100 ages from 10--300 Myr sampled uniformly in log-space for young field targets, and 30 ages from 5--15 Myr sampled uniformly in log-space for Sco-Cen targets. We then used the python package \texttt{ExoDMC} \citep{Bonavita2020} to generate a detection sensitivity map for each sampled age for each target. \texttt{ExoDMC} is a Monte Carlo simulation code, which generates a synthetic population of 1000 planets with a range of orbital parameters. The assumptions and models used to generate this population are explained in \cite{Bonavita2012}. We combined the detection sensitivity maps generated for each star at each age to estimate our overall detection sensitivity for young field stars (this includes both the LASSO and pre-LASSO stars, as their sensitivity maps were nearly identical) and Sco-Cen association stars, shown in Figure \ref{fig:detection_sensitivity} and summarized in Table \ref{tab:detection_sensitivity}.}

\begin{table}[h]
\centering
\caption{\new{Robo-AO Survey Detection Sensitivity}
    \label{tab:detection_sensitivity}}
\begin{tabular}{cccc}
\new{\textbf{Companion}} & \multicolumn{3}{c}{\new{\textbf{Detection probability}}} \\
\new{\textbf{Mass}} & \new{\textbf{10\%}} & \new{\textbf{50\%}} & \new{\textbf{90\%}} \\
\hline
\multicolumn{4}{c}{\new{LASSO field stars}} \\
\hline
\new{70 M$_{Jup}$} & \new{10--1425 AU} & \new{50--450 AU} & \new{$\cdots$} \\
\new{40 M$_{Jup}$} & \new{15--1100 AU} & \new{95--300 AU} & \new{$\cdots$} \\
\new{13 M$_{Jup}$} & \new{45--440 AU} & \new{$\cdots$} & \new{$\cdots$} \\
\hline
\multicolumn{4}{c}{\new{Sco-Cen association stars}} \\
\hline
\new{70 M$_{Jup}$} & \new{15--2000 AU} & \new{45--1050 AU} & \new{210--570 AU} \\
\new{40 M$_{Jup}$} & \new{30--2000 AU} & \new{80--1045 AU} & \new{330--525 AU} \\
\new{13 M$_{Jup}$} & \new{105--1525 AU} & \new{450--500 AU} & \new{$\cdots$} \\
\hline
\end{tabular}
\end{table}

\subsubsection{Visible camera}

The Robo-AO automated data reduction pipeline for the visible camera produces PSF-subtracted images and contrast curves for each observation. We visually inspected the images to flag companion candidates. Companion candidate measurements and SNR values were calculated in the same way as described for the infrared camera above. We used the contrast curves generated by the high contrast pipeline for each observation to determine the sensitivity of our survey in the visible. Details about the performance and achievable contrasts of Robo-AO in the visible can be found in Figure 13 of \cite{JensenClem18}.

\subsubsection{False Triples}
For some bright companions, the frame stacking may lock onto the centroid of the companion instead of the primary in some of the frames. This produces a ``false triple" image where the companion appears on both sides of the primary, at the same separation and 180$^{\circ}$ rotated. We measured the contrast for both locations of the companion, masking both companion locations when measuring the background and noise in the annulus around the star. We then combined the measurements following \cite{Law06} to get the final contrast. This effect can be seen in both the visible and infrared data pipelines. We corrected four of our companion candidate contrast measurements for this effect.

\section{Results} \label{sec:results}

We detected a total of \Ncomps companion candidates near \Nhostcandidates stars using the Robo-AO infrared camera: \NcompsKP in the \textit{J}-band at Kitt Peak and \NcompsUH in the \textit{H}-band at Maunakea, (including \NhostsKPUH observed and detected at both telescopes). We acquired simultaneous images with the visible camera of \NhostsVIC stars with \NcompsVIC companion candidates and detected \NcompsVICdet of these companion candidates \new{in the $i'$-band}. The resulting \new{detection} measurements are summarized in Table~\ref{tab:comps_measurements} in the Appendix.

\subsection{Physical association}
\label{sec:bound}

\begin{figure}[ht!]
\centering 
\includegraphics[width=240pt]{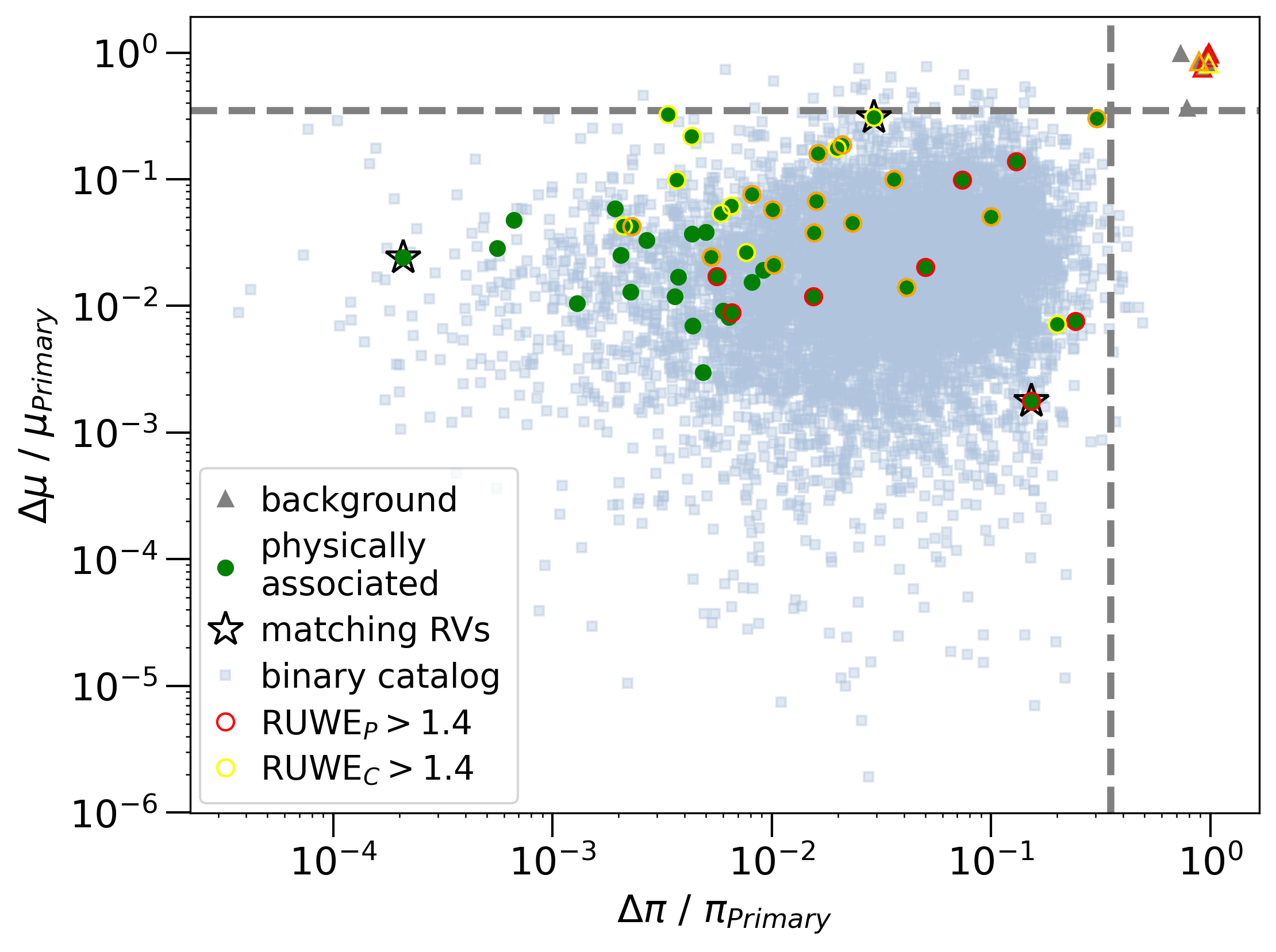}
\caption{Comparing \textit{Gaia} DR2 parallaxes and proper motions for primary stars and companion candidates to determine consistency for physical association. For reference, the co-moving binaries identified in the \textit{Gaia} DR2 binary catalog by \cite{Jimenez19} are also shown. For objects to be considered physically associated, we set the threshold for the ratios of primary-companion parallax difference to primary parallax ($\Delta \pi/\pi_{Primary}$) and primary-companion proper motion difference to primary proper motion ($\Delta \mu/\mu_{Primary}$) $<$ 0.35 (dashed lines). In addition, we have \textit{Gaia} DR2 RVs for three primary-companion pairs (star symbols), which all match within $\Delta RV < 2 \sigma_{RV}$. Pairs with less reliable astrometry (RUWE $>$ 1.4) are marked in colored circles: red if the primary star's RUWE $>$ 1.4, yellow if the companion's RUWE $>$ 1.4, and orange if both RUWEs $>$ 1.4.
    \label{fig:prlx_pm_match}}
\end{figure}

\begin{figure}[ht!]
\centering
\includegraphics[width=240pt]{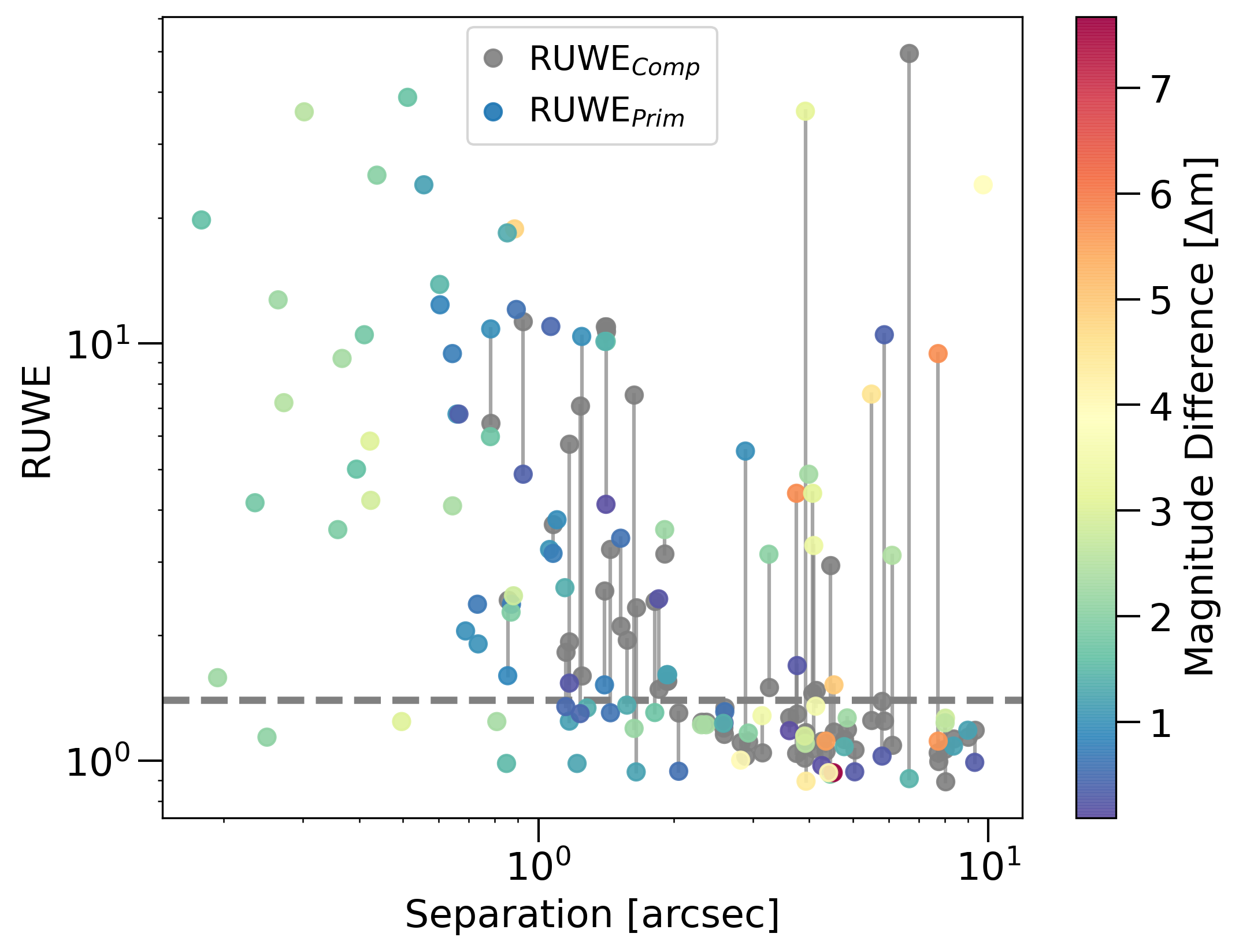}
\caption{RUWE values for the primary (colored by magnitude difference) and companion (grey, connected by a grey line to its primary) as a function of separation. A trend of large RUWE values ($>1.4$, above horizontal grey dashed line) for closer in companions is visible, however no particular trend with contrast is obvious.
    \label{fig:RUWE_sep}}
\end{figure}

We searched \textit{Gaia} DR2 for objects at the same location as our companion candidates and found matches for \NcompsGaia of them. Of those, \Ncompsprlxs had parallax and proper motion measurements. To estimate the possible physical association of the companion candidate and primary star, we compared their parallaxes and proper motions. We calculated the ratios of the primary-companion difference in parallax ($\Delta \pi$) to the parallax of the primary ($\pi_{Primary}$) and the primary-companion difference in proper motion ($\Delta \mu$) to the proper motion of the primary ($\mu_{Primary}$). In order to establish reasonable thresholds on these ratios, we compared our sample of primary-companion pairs to the \textit{Gaia} DR2 selected wide co-moving binaries from \cite{Jimenez19}, shown in Figure \ref{fig:prlx_pm_match}. \cite{Jimenez19} determine pairs to be comoving if their differences in parallax and proper motions in RA and Dec are less than $2 \sigma$, where $\sigma$ is the maximum error of the two measurements. However, we opted to not directly include the errors in our thresholds because the majority of our candidates have large ($>1.4$) Renormalized Unit Weight Errors (RUWE) values, which indicates an issue with the astrometry solution, which uses a single star model. Such a large RUWE could indicate the \textit{Gaia} DR2 measurements are affected by the presence of the companions. \cite{Jimenez19} only included sources with RUWEs $<1.4$ and made further sample restrictions to ensure reliable error estimates. Figure \ref{fig:RUWE_sep} shows RUWE values for our sample as a function of separation. Stars with closer companions have large RUWEs, which suggests that the large RUWE is due to the presence of a companion. Therefore, we have determined that objects with $\Delta \pi/\pi_{Primary}$ or $\Delta \mu/\mu_{Primary}$ $>$ 0.35 to very likely be background objects. 

Twelve objects clearly stand out as background objects, while the other \NcompsGaiabound companions all appear to be consistent with physical association. In addition, three of our pairs have \textit{Gaia} DR2 radial velocity (RV) measurements for both components, which are in agreement ($\Delta RV < 2 \sigma_{RV}$, where $\sigma_{RV}$ is the maximum error of the two measurements), and thus further indicates the likelihood of physical association. \new{Of the remaining} \NcompsGaiaunknown companion candidates lacking \textit{Gaia} DR2 measurements, \new{\NcompsLitbound were previously studied systems reported in the literature as physically associated companions. Another five have astrometry measurements in Gaia EDR3\footnote{\new{Gaia EDR3 was released after this paper was submitted. We have used this to update the physical association status of five companion candidates, which were previously unconfirmed with Gaia DR2. We also confirmed that the status of the other candidates remained unchanged. Except for these five companions, the properties reported throughout the paper are calculated using distances from Gaia DR2 astrometry.}}, three of which are physically associated companions and two are background objects. Therefore, a total of \Ncompsbound companions are physically associated, \Ncompsbackground are background objects, and} \Ncompsunknown candidates will require future follow-up observations \new{to asses their physical association}. 

\begin{figure}[ht!]
\centering
\includegraphics[width=240pt]{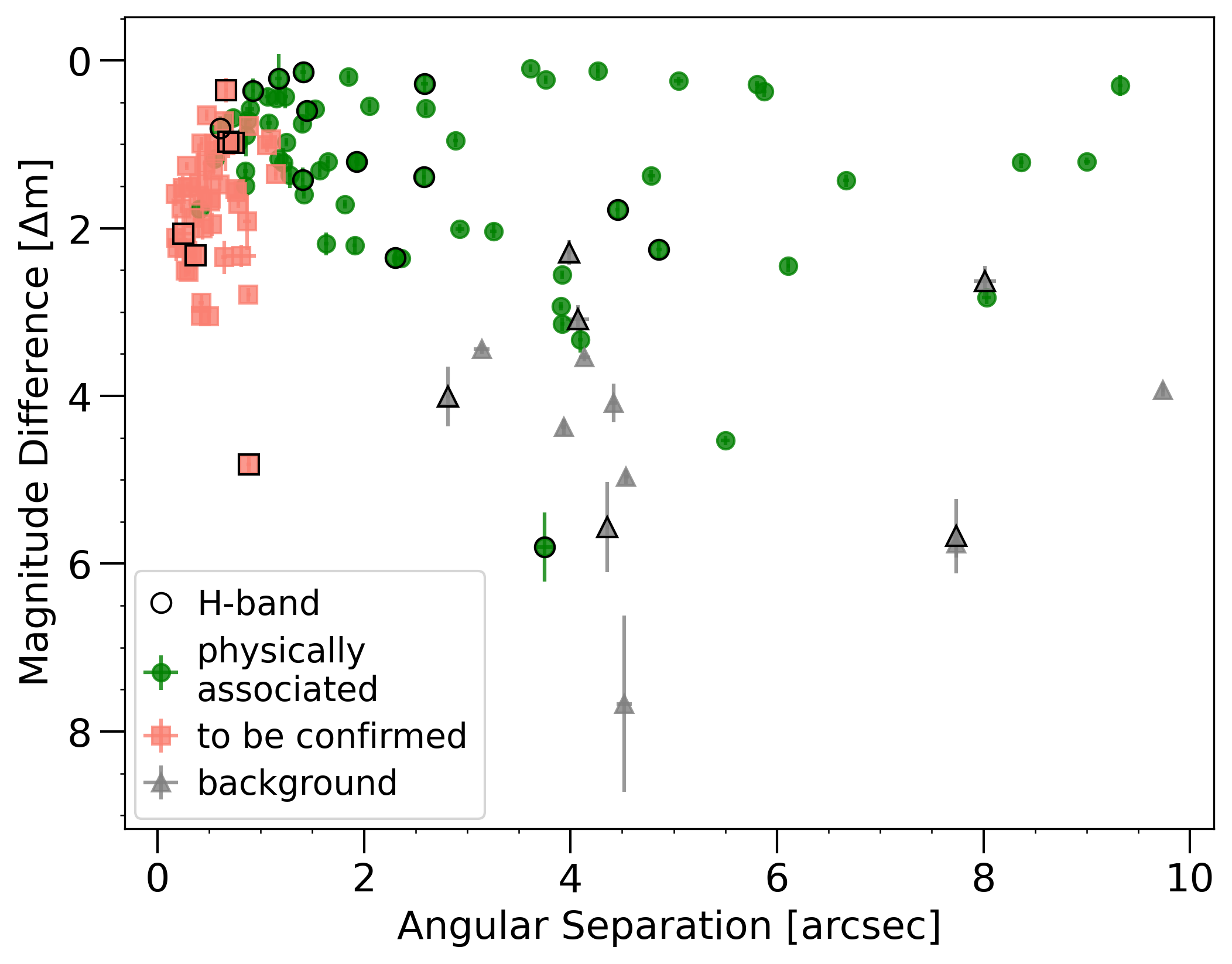}
\caption{Summary of companion candidates detected with Robo-AO as part of this survey. Companions with \textit{Gaia}-DR2 parallaxes and proper motions consistent with their primary stars are determined to be physically associated (green circles). Companions not in \textit{Gaia}-DR2 or without parallax or proper motion information still need their physical association status to be determined (salmon squares). Companion candidates with parallax and proper motions inconsistent with the primary star were determined to be background objects (grey triangles). Measurements from Kitt Peak are in the \textit{J}-band and measurements from Maunakea are in the \textit{H}-band (markers outlined in black).
    \label{fig:comps_arcsec_dmag}}
\end{figure}

\begin{figure}[ht!]
\centering
\includegraphics[width=240pt]{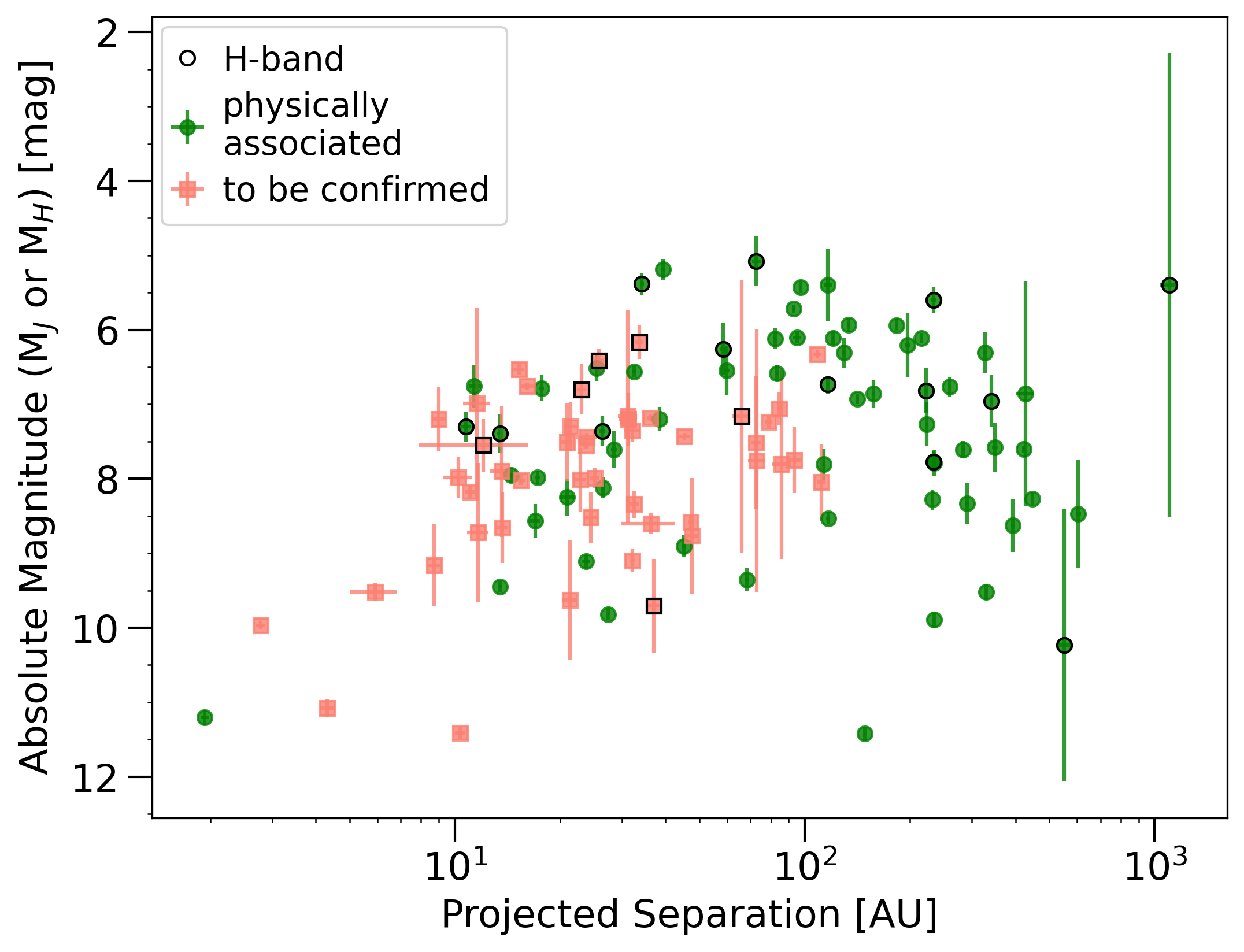}
\caption{Absolute magnitudes and projected physical separations of companion candidates. Color schemes are the same as Figure \ref{fig:comps_arcsec_dmag}. Candidates without enough information to determine physical association were assumed to be at the same distance as their host stars for the calculations. The majority of the unconfirmed candidates are within 100 AU.
    \label{fig:comps_AU_absMag}}
\end{figure}

\begin{figure*}[ht!]
\centering
\includegraphics[width=495pt]{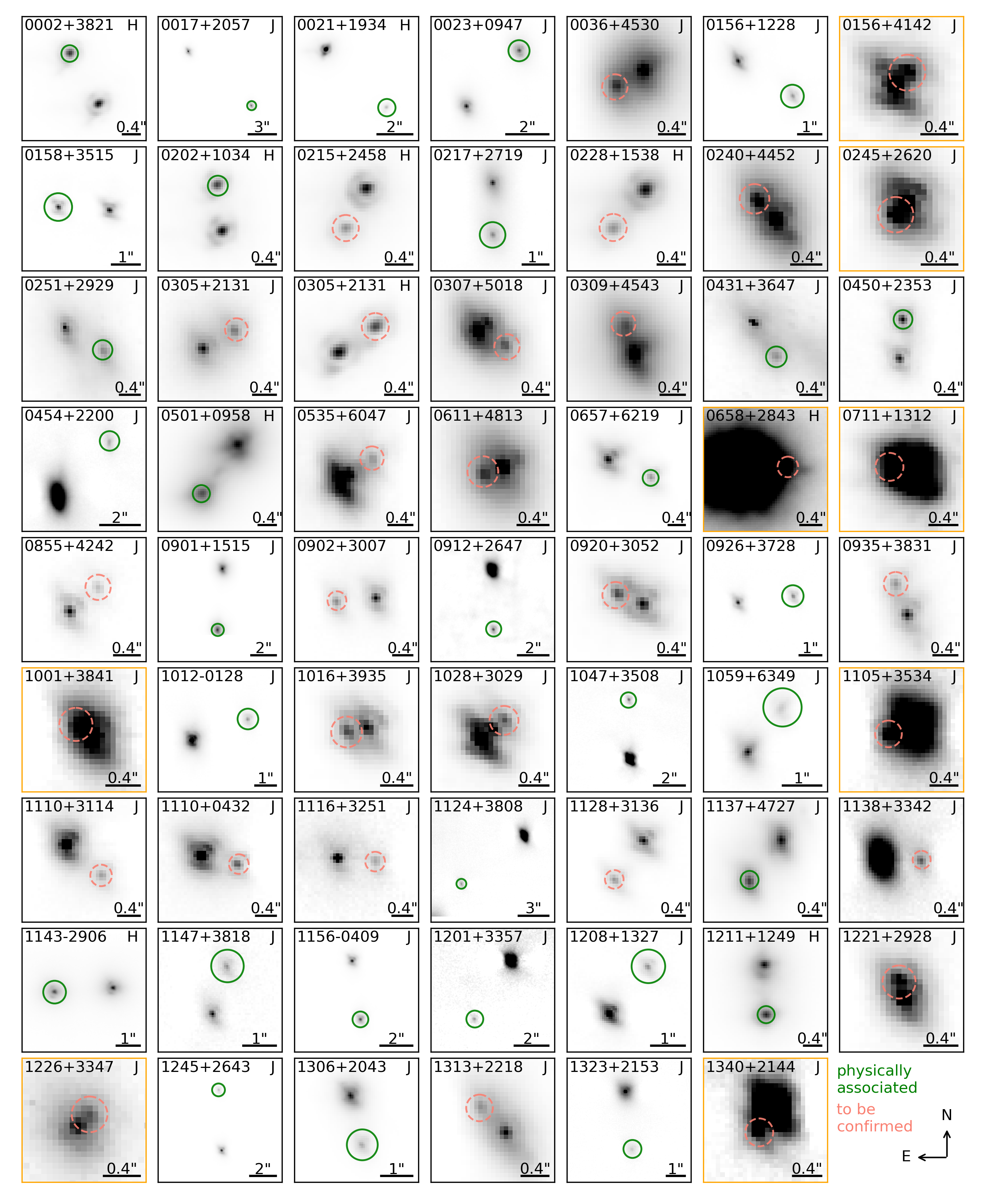}
\caption{Linear scale images of Robo-AO near-infrared binary detections of physically associated companions (green circles) and unconfirmed candidates (dashed salmon circles) with the dynamic range adjusted for companion visibility. The filter used for the observation is shown in the top right and the image scale is shown in the bottom right. Images outlined in orange are shown in Figure \ref{fig:img_grid2}b. Triple systems are shown in Figure \ref{fig:triple_img_grid}.
    \label{fig:img_grid1}}
\end{figure*}

\begin{figure*}[ht!]
\subfloat{(a)}{\includegraphics[width=495pt]{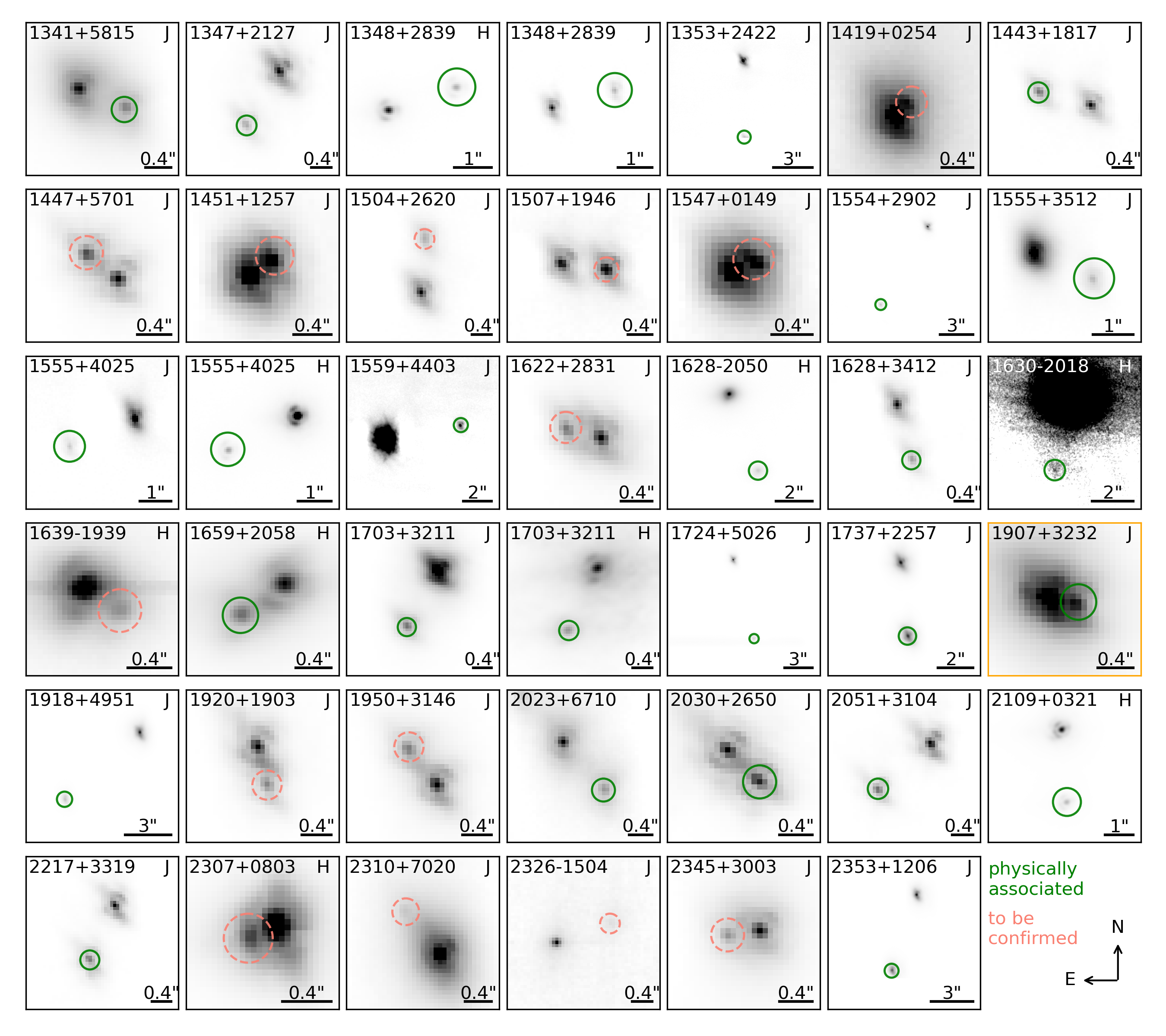}}\\
\subfloat{(b)}{\includegraphics[width=495pt]{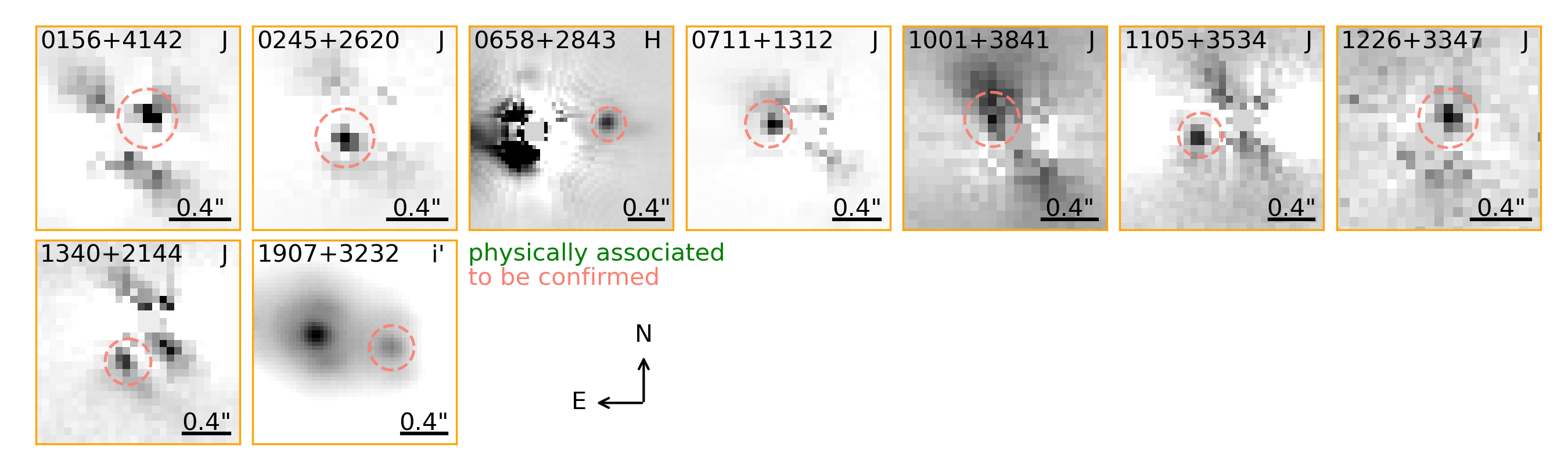}}
\caption{\new{(a)} Linear scale images of Robo-AO near-infrared binary detections of physically associated companions (green circles) and unconfirmed candidates (dashed salmon circles) with the dynamic range adjusted for companion visibility. The filter used for the observation is shown in the top right and the image scale is shown in the bottom right. Images outlined in orange are shown in \new{(b)}. Triple systems are shown in Figure \ref{fig:triple_img_grid}. \new{(b)} Radially subtracted images for very close companion candidates not easily seen in Figures \ref{fig:img_grid1} \& \ref{fig:img_grid2}a. Except for 2MASS J19074283+3232396 where it is most easily seen in the \textit{i'}-band image.
    \label{fig:img_grid2}}
\end{figure*}

We show the companion infrared contrasts \new{and absolute magnitudes} as a function of separation from the host star, with physical association status, in Figures~\ref{fig:comps_arcsec_dmag} \new{\& \ref{fig:comps_AU_absMag}.} A summary of the physically associated companions in binary systems and their properties are shown in Appendix Table \ref{tab:comp_properties_binaries_bound}. Companion candidates where physical association has yet to be confirmed are summarized in Appendix Table \ref{tab:comp_properties_binaries_unknown}. \new{We display images of all of the confirmed and unconfirmed binary companions in Figures \ref{fig:img_grid1} and \ref{fig:img_grid2}.} For now, we assume the same parallax as the primary star to estimate their properties. Triple system candidates are summarized in Appendix Table \ref{tab:comp_properties_triples} and further discussed in the section \ref{sec:triples}. As seen in both figures, most of the companions that do not have \textit{Gaia} DR2 information are near the angular resolution limit of \textit{Gaia} where accurate astrometry is more challenging. \cite{Hirsch17} and \cite{Horch14} have shown that the probability of a background star chance alignment decreases as the separation on-sky decreases. Therefore, we expect the majority of these close-in companion candidates to be physically associated.

\new{We counted the stars surrounding each target with an unconfirmed companion candidate that were brighter than the faintest magnitude reached in all of the contrast curves ($<$17 in $J$-band and $<$18 in $H$-band). We conducted the search using the 2MASS catalog in a circular area of 20$\arcmin$ radius. We extrapolated the star count in each band beyond the 2MASS 10$\sigma$ sensitivity limits of 15.9 and 15.0 in $J$ and $H$ bands, respectively. The UKIDSS catalog \citep{Lawrence07} reaches fainter IR magnitudes, avoiding the need to extrapolate. However it does not cover the entire sky, and none of our targets with unconfirmed companion candidates were found in UKIDSS. We then divided the count by the search area in order to get the stellar surface density, then multiplied by the Robo-AO IR field of view to get the expected cumulative star counts within that area ($N_{FOV}$). The probability of detecting at least one background object was then calculated assuming a Poisson distribution:} 
\begin{equation}
    \new{P(N_{background}\geq1) = 1 - e^{-N_{FOV}}}
\end{equation}
\new{The resulting probabilities and their Poisson errors are reported in Tables  \ref{tab:comp_properties_binaries_unknown} \& \ref{tab:comp_properties_triples}. Only five of the \Ncompsunknown companion candidates have $>$10\% chance of a background star landing in the field of view of our observation. Of those, three are below 20\%, and the remaining two values are both 43\%. This reinforces our expectation that the vast majority of our companion candidates are true companions. However, it is important to note that although each individual companion candidate has a low probability of being a background object, this is not enough to claim that any specific companion candidate is not a background object. If we calculate the compound probability that at least one background object is detected around any of the \Ncompsunknown targets, we get a 93\% probability.} 

\subsection{Optical - Infrared Colors}
\label{sec:colors}

\begin{figure*}[ht!]
\begin{center}
\subfloat{\includegraphics[width=245pt]{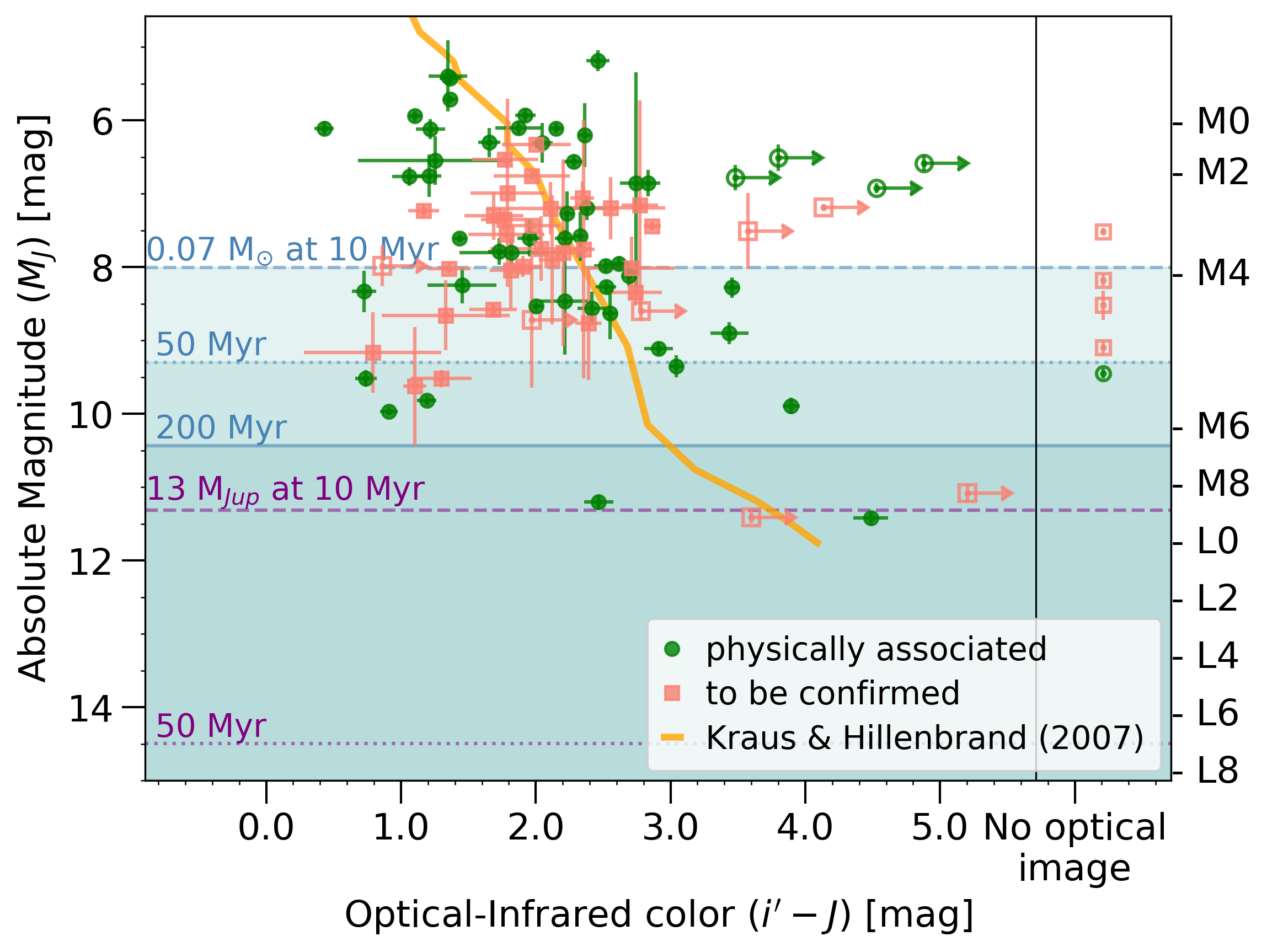}}
\subfloat{\includegraphics[width=245pt]{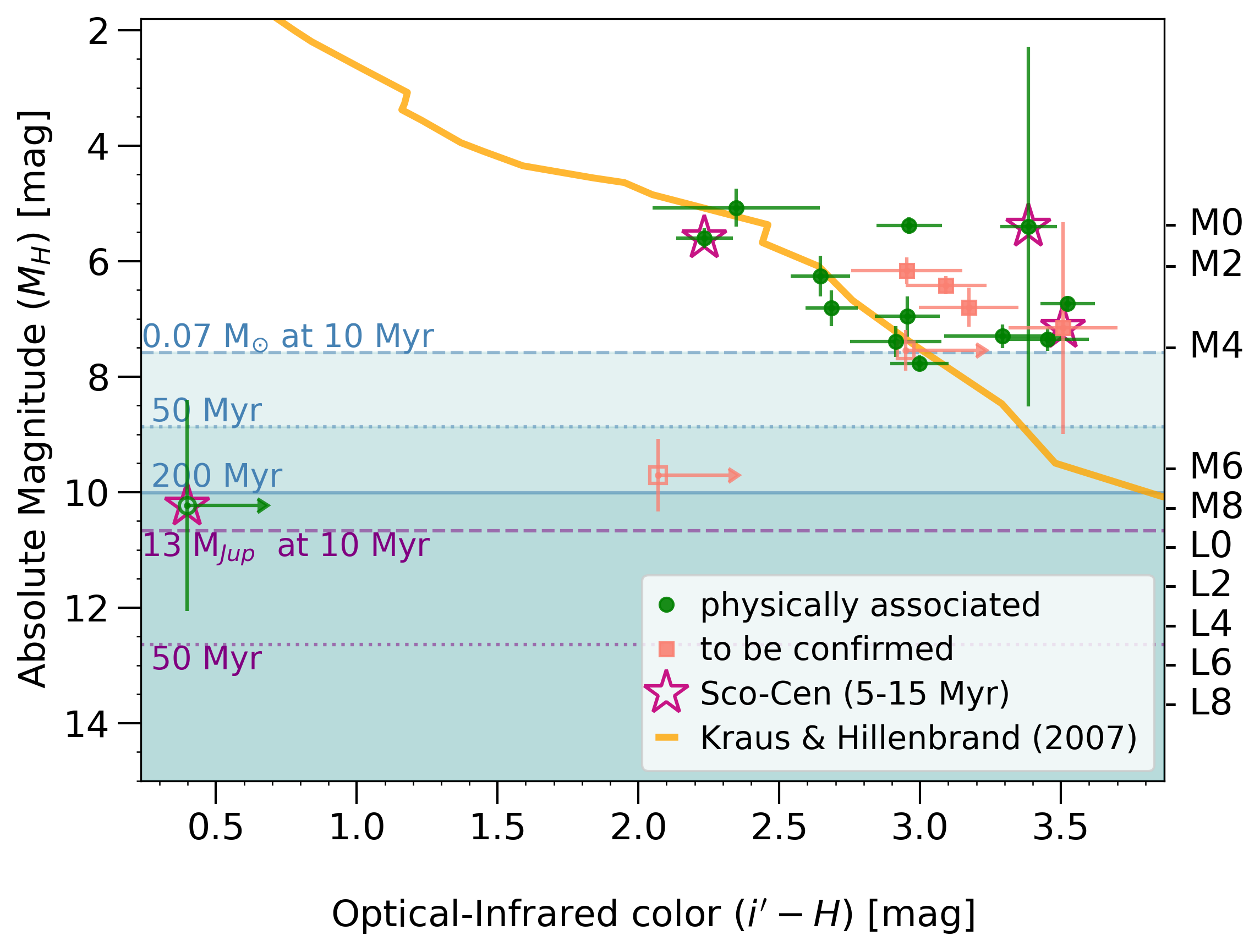}}
\caption{Companion candidate absolute \textit{J} (\textit{left}) and \textit{H} (\textit{right}) magnitudes as a function of optical-infrared colors. Targets with a companion candidate detected in the infrared but not in the visible are shown with lower limits on their colors, denoted by open symbols with arrows. Companions without \textit{i'}-band photometry are shown to the right of the solid black vertical line. Companions that are physically associated are shown in green circles, while those that still need to be confirmed are shown as salmon squares. Corresponding mass estimates from isochrones of 10, 50, and 200 Myr are shown to delineate the stellar-substellar boundary (shaded blue area) and the deuterium burning limit (purple dashed and dotted lines). Companion candidates in the Sco-Cen sample are outlined with a pink star. We estimated the spectral types using the stellar SEDs in \cite{Kraus07} for types earlier than M6 and  \cite{Liu16}'s linear relation for M6 -- L8 field objects.
    \label{fig:comps_absmags_VisIRcols}}
\end{center}
\end{figure*}

For stars with simultaneous visible and infrared images, we report \textit{i'}-\textit{J} or \textit{i'}-\textit{H} colors for the companion candidates. For the \NcompsNOVICdet objects not detected in the visible images, we placed lower limits on their colors from the visible contrast limits. For companion candidates where physical association has yet to be confirmed, we used these color measurements combined with absolute magnitudes, and assuming the companion is at the same distance as the primary, to determine whether their photometry is consistent with a low-mass companion or background star. We used evolutionary models \citep{Chabrier00,Baraffe15} to estimate companion masses and temperatures from absolute magnitudes. In Figure \ref{fig:comps_absmags_VisIRcols}, we show the companion candidates' absolute magnitudes as a function \neww{of} optical-infrared colors with their physical association status. \neww{We also show spectral type and mass estimates from stellar color-magnitude sequences \citep{Kraus07,Liu16} and isochrone models \citep{Chabrier00,Baraffe15}. Companion candidates that are potentially substellar will be prioritized for follow-up observations (see \S \ref{sec:substellar}).}

\subsection{Literature Search for Companions}

We found \Ncompscat of our \Ncomps ~companion candidates in previous catalogs, listed in Appendix Table \ref{tb:catalog_matches}. The following sections summarize the results.

\subsubsection{Robo-AO M dwarf multiplicity survey} 
\cite{Lamman20} surveyed 5566 field M dwarfs at visible wavelengths with Robo-AO to assess any multiplicity. They found 553 companion candidates within 4$\arcsec$ of 534 different stars. Seven of our companion candidates are also reported in this catalog.

\subsubsection{Imaging of CARMENES M dwarfs}
\cite{CortesContreras17} searched for low-mass companions to M dwarfs to vet targets for the CARMENES exoplanet survey. They observed 490 stars, from a volume-limited sample of M0-M5 stars within 14 pc. They found 80 bound companions and six companion candidates. We detected three of their confirmed companions. Due to a lack of \textit{Gaia} DR2 measurements for two of those companions, the CARMENES input catalog was used to determine their physical association.

\subsubsection{Young binaries and lithium-rich stars}
\cite{Bowler19} searched for new young, nearby, low-mass stars and report on spectroscopic observations of lithium-rich stars and binaries identified with Robo-AO. Nine of our companion candidates were identified with Robo-AO in this study, including two (2MASS J12115308+1249135 \& 2MASS J15553178+3512028) identified as members of young moving groups $\beta$ Pic and Argus, respectively.

\subsubsection{Washington Double Star catalog}
We searched for our companions in the Washington Double Star (WDS) catalog \citep{Mason01}. We found 27 of our companion candidates in the WDS, including four with unconfirmed physical association and one that we ruled out as a background object (2MASS J02022823+1034533) but is listed in the WDS.

\section{Discussion} \label{sec:Discussion}

\subsection{Triple Systems}
\label{sec:triples}

We discovered \new{\Ntriples} triple system candidates; \new{six} of which include an object determined not to be physically associated from \textit{Gaia} DR2 parallaxes and proper motions. Of the remaining \new{\NtriplesNOTbkg} triple system candidates, \new{three} are known triples. The remaining system \new{is} potentially \new{a} new triple system, requiring follow-up observations to confirm. 


The presence of a tertiary companion and resulting architecture of the system gives important insight on the formation and evolution of high multiplicity systems. Hierarchical systems in particular, with a tight secondary and wider tertiary companion, have been shown to be the structure reached as orbits stabilize \citep{Reipurth12}. Furthermore, the structure of our triple system candidates can help us infer which ones are more or less likely to be a true triple system with both companions physically associated. Specifically, those that are not found in a hierarchical system are less likely to be true triple systems, because a hierarchical structure is the only stable structure. 

\begin{figure}[h!]
\centering
\includegraphics[width=245pt]{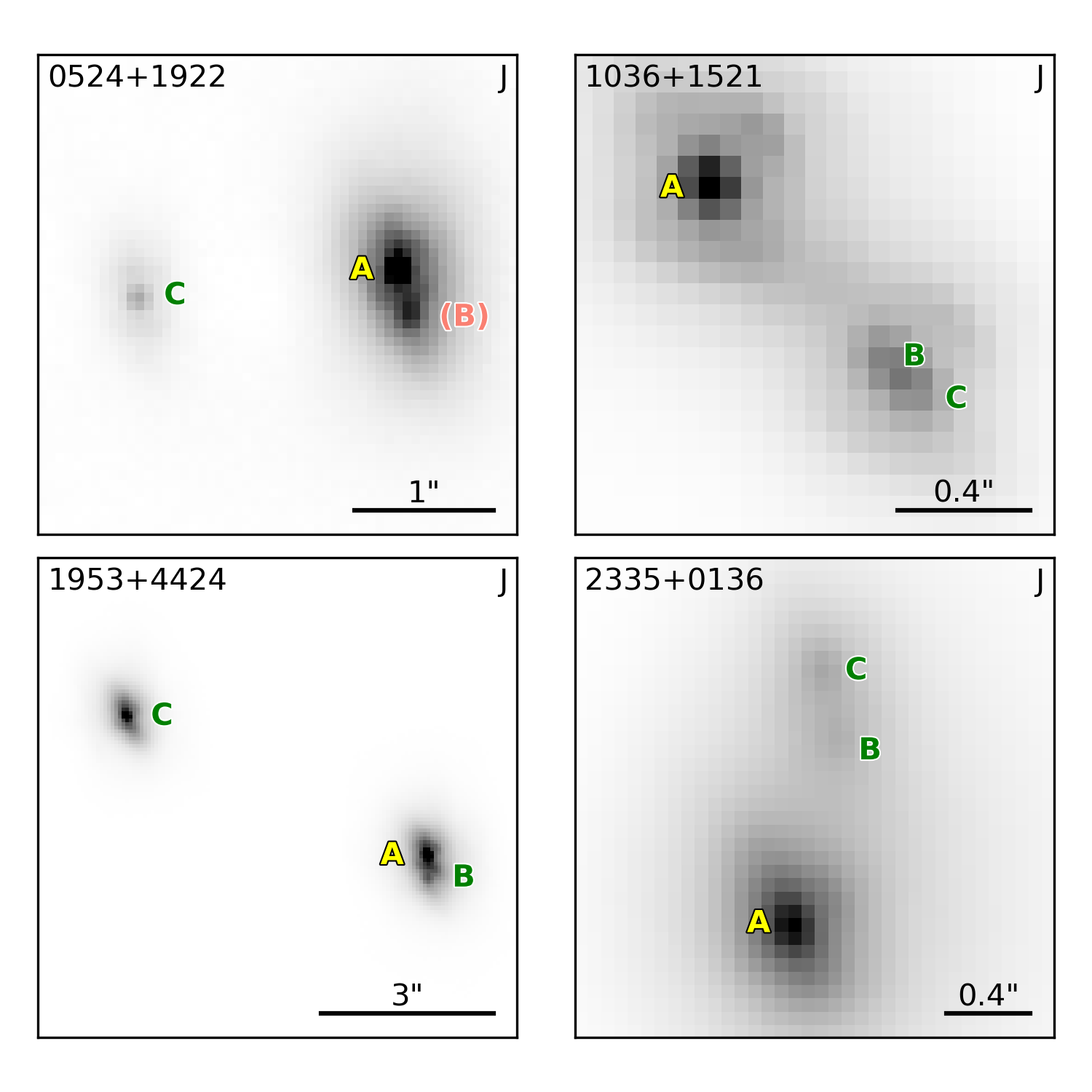}
\caption{Images of the triple system candidates. The companions that were determined to be physically associated, either from \textit{Gaia} DR2 astrometry or from the literature, are labeled `B' or `C' in green, while \new{the one} whose physical association has yet to be confirmed \new{is} labeled `(B)' in salmon. The primary star is labeled with a yellow `A'. The 2MASS ID and the Robo-AO IR filter are shown in the upper left and right, respectively. The images are displayed in linear stretch with the dynamic range adjusted for the faintest companion visibility. \new{North is up and East is left.}
    \label{fig:triple_img_grid}}
\end{figure}

Images of the \new{\NtriplesNOTbkg} triple system candidates are shown in Figure \ref{fig:triple_img_grid} \new{and each system is detailed below, the known systems first followed by the new candidate system}.

2MASS J10364483+1521394 is an M4.5 rotational variable star with flares \citep{Rodriguez20} and is a triple system at a distance of 19.75~pc. We detect the companions at projected separations of 14.82~AU and 17.78~AU, with the companions forming a nearly equal-mass binary at a projected separation of 0.15$\arcsec$ (3~AU). \cite{Calissendorff17} recently analyzed the orbits and masses of this known triple system.

2MASS J19535443+4424541 is a known hierarchical triple system \citep{Tokovinin17}, with a tight central binary and a farther out fainter companion. The primary is an M5.5Ve star, in a tight binary with a $\sim$0.1M$_{\odot}$ companion and the wider companion is an M6V star. This is one of the nearest triple systems, at a distance of 4.7 pc. The inner binary is separated by 1.9 AU and the outer companion is at a separation of 27.2 AU. \cite{Tokovinin17} determined the inner binary to have a period of 15.2 years and eccentricity of 0.32 and the outer companion's orbit direction is retrograde.

2MASS~J23350028+0136193 is a K7V star at a distance of 18.2~pc and a member of the IC~2391 moving group with an age estimate of $50\pm5$ Myrs \citep{Faherty18,Barrado04}. This is a triple system, with recent measurements on the fainter and closer companion reported in \cite{Mann19} and \cite{Kammerer19}. The projected separation of the companions are 17.70 and 25.37~AU. \cite{Mann19} reports a total system mass of $0.606\pm0.018 M_{\odot}$. \cite{Kammerer19} report an RV of 4.5 km/s in the HARPS RV survey. 

\new{2MASS J05242572+1922070 is a weak-line T Tauri star \citep{Li98} and a disk-free member of the Taurus-Auriga star-forming complex \citep{Kraus17} at a \textit{Gaia} DR2 distance of 58.34$\pm$1.01 pc. The outer companion is at a projected separation of 113.2$\pm$1.3 AU, and we have determined it to be physically associated from \textit{Gaia} DR2 proper motion and parallax measurements. Using isochrone models, we estimate its mass to be 84--319 M$_{Jup}$ for an age range of 10--200 Myrs. The inner companion candidate is at a projected separation of 20.9$\pm$0.7 AU and remains to be confirmed. However, we expect this system to be a true new triple system given the overall hierarchical appearance of the system, the companion candidate's very close separation (0.36$\arcsec$), the primary star's RUWE value $>$1.4 (3.6 in \textit{Gaia} DR2 and 1.8 in \textit{Gaia} EDR3), and the low probability of a chance alignment with a background star (7.95$\pm$0.13\%).}

\subsection{Accelerating Stars}
\label{sec:accelerations}

\begin{figure}[ht!]
\centering
\includegraphics[width=240pt]{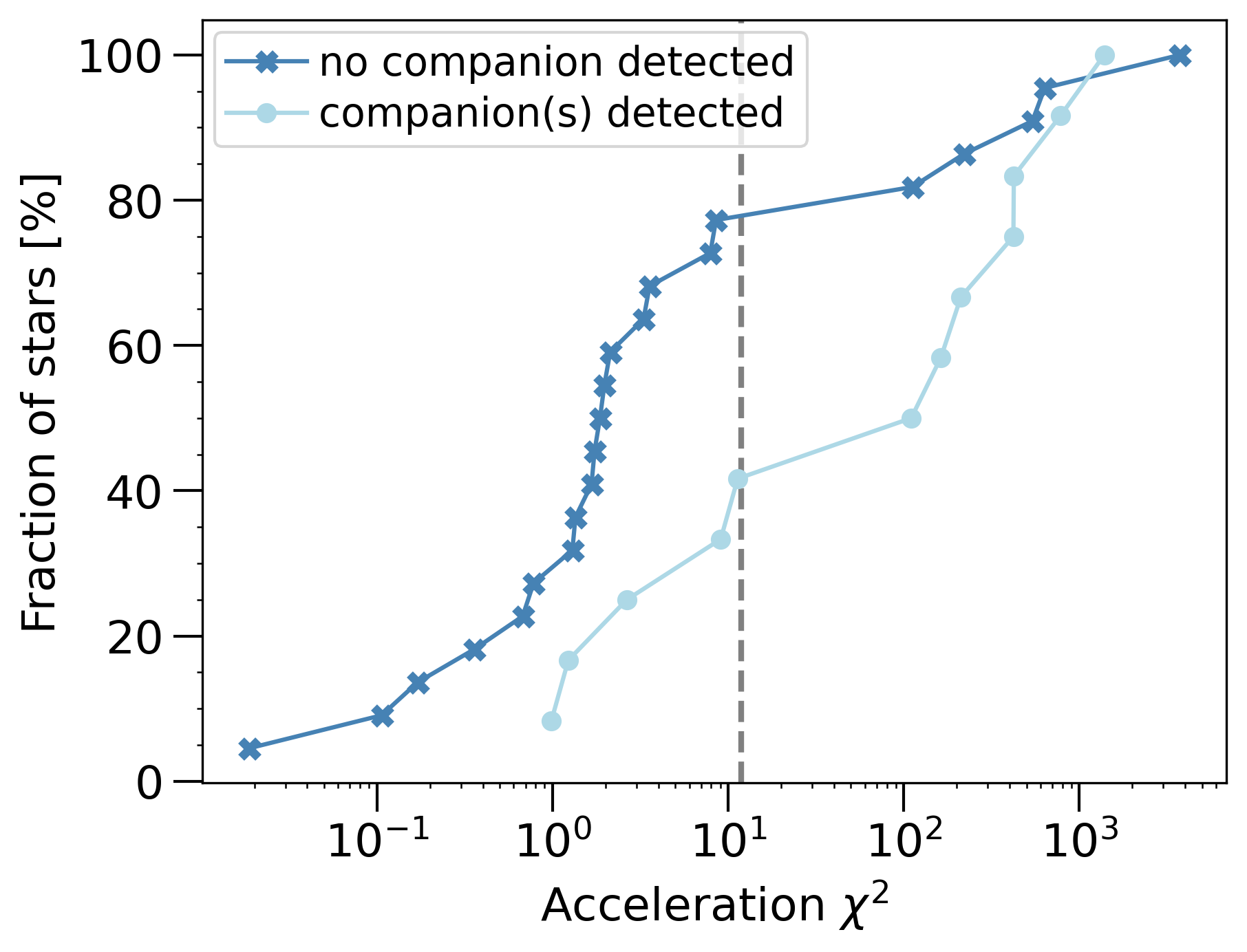}
\caption{\new{Cumulative distribution} of acceleration $\chi^2$ value\new{s} for stars in the HGCA catalog \new{comparing} stars with no Robo-AO companion \new{detection}s \new{to} those with companion candidates. Stars with $\chi^2 > 11.8$ are accelerating with $>3\sigma$ significance. A larger proportion\new{, 58$^{+12}_{-14}$\% (7/12),} of stars with companion detections are accelerating than those with no companion detection\new{, 23 $^{+11}_{-6}$\% (5/22)}.
    \label{fig:comps_accelerations_hist}}
\end{figure}

The \textit{Hipparcos-Gaia} Catalog of Accelerations (HGCA; \citealt{Brandt18}) measured accelerations by using three proper motion and positional measurements from \textit{Hipparcos} (near epoch 1991.25), \textit{Gaia} DR2 (near epoch 2015.5), and the \textit{Gaia} -- \textit{Hipparcos} scaled positional difference over the 24-year baseline. Stars with measured accelerations are particularly interesting as they can provide dynamical masses and orbits for companions \citep{Brandt19}. We consider stars with $\chi^2 > 11.8$, corresponding to 3$\sigma$, calculated from the \textit{Gaia} proper motions against the \textit{Gaia} -- \textit{Hipparcos}, to have significant accelerations. Of our observed targets, 34 are found in the HGCA catalog\new{: 22 with no imaged companion and} twelve with imaged companion candidate(s). The \new{cumulative} distribution of acceleration $\chi^2$ are shown in Figure \ref{fig:comps_accelerations_hist}. 

\new{Of the observed stars found in the HGCA catalog where we did not detect a companion, 77$^{+6}_{-11}$\% (17/22)} do not have significant accelerations ($\chi^2 < 11.8$), and \new{23$^{+11}_{-6}$\% (5/22)} have significant accelerations. \new{Of those five stars, two are reported in the literature as having a tight companion: 2MASS J13232325+5754222 \citep{Horch17} and 2MASS J22372987+3922519 \citep{Pourbaix04}. It is possible that there are also unresolved companions around the remaining three stars with significant acceleration measurements but where we do not detect any companions.}

\begin{figure}[ht!]
\centering
\includegraphics[width=240pt]{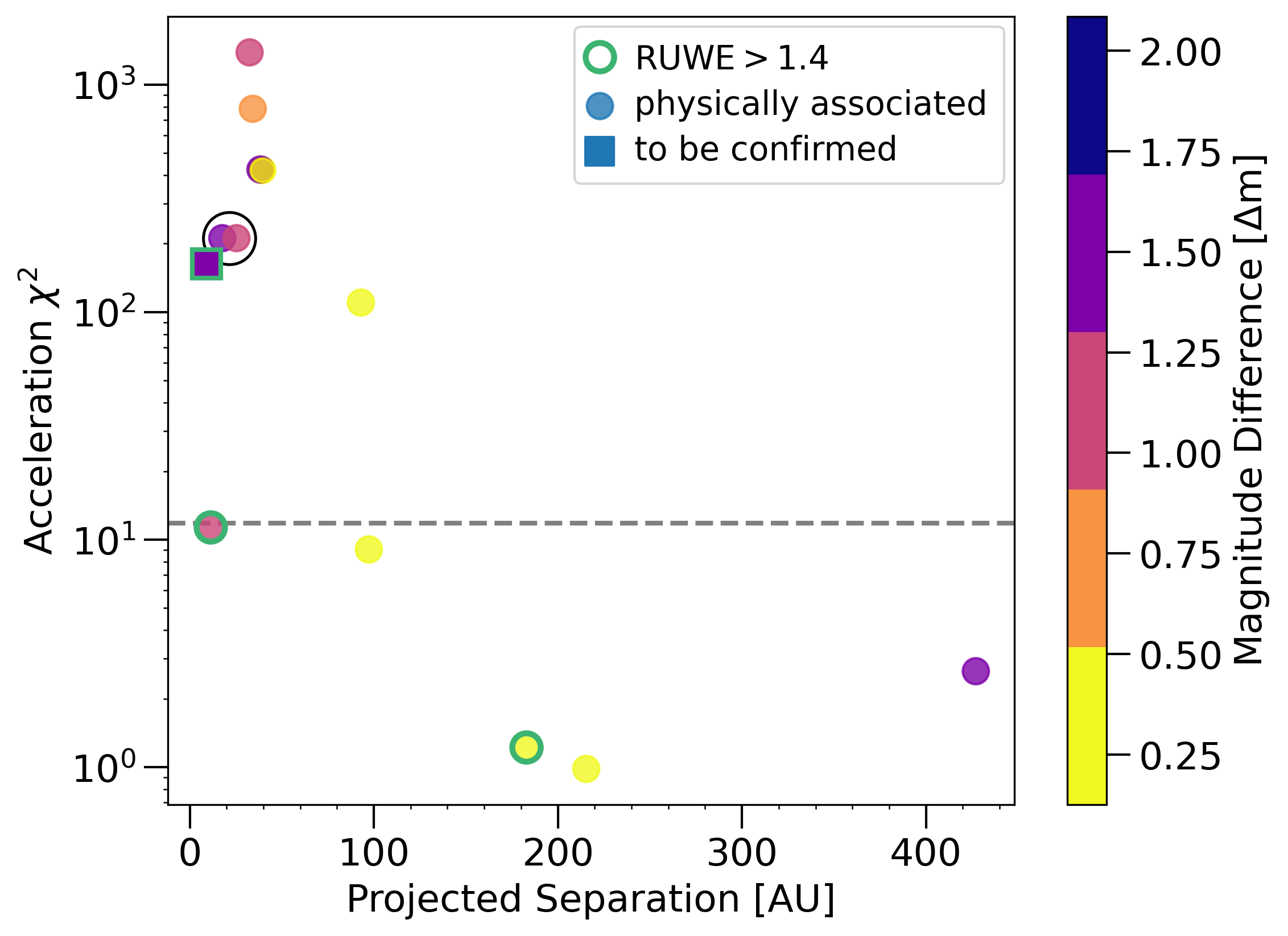}
\caption{Target acceleration significance from the HGCA catalog as a function of companion projected separation. The threshold determining ``significant acceleration" is placed at $\chi^2 > 11.8$ (dashed grey line) and targets with large RUWE values, and thus less reliable astrometry, are outlined in green. Targets with companions that have been confirmed to be physically associated are marked as circles and \new{the one} that still need\new{s} to be confirmed \new{is} marked as \new{a} square, but \new{is} highly likely to be physically associated due to \new{its} high acceleration significance. One system is a triple system, and its two companions are circled in black. A correlation is apparent where closer companions have large acceleration sign\new{i}ficances.
    \label{fig:comps_accelerations}}
\end{figure}

In contrast, for \new{the stars} with companion detection(s), \new{58$^{+12}_{-14}$\% (7/12)} have significant accelerations and \new{42$^{+14}_{-12}$\% (5/12)} do not. \new{The 1-$\sigma$ uncertainties are numerically calculated following the binomial distribution, as described in \cite{Burgasser03}.} Figure \ref{fig:comps_accelerations} shows that there is a clear correlation between acceleration significance and companion projected separation, as expected. Similarly, the companion that do\new{es} not have \textit{Gaia} DR2 measurements but ha\new{s} significant acceleration (2MASS J03093085+4543586) \new{is} very likely to be \new{a} physically associated companion.

\subsection{Substellar Candidates}
\label{sec:substellar}

While the majority of our detections are very likely stellar companions, a handful are potentially substellar. We do not have precise age estimates for most of our targets; therefore these systems will require additional infrared color photometry and spectroscopy to characterize these objects and determine their spectral types. Table \ref{tb:known_comps} summarizes the companions we detected with Robo-AO in LASSO that have already been discovered and characterized but would have been flagged as potentially substellar and needing follow-up study from the Robo-AO data.

\begin{table}[h]
\centering
\caption{LASSO companions found in the literature
    \label{tb:known_comps}}
\begin{tabular}{lcccl}
\textbf{2MASS ID} & \multicolumn{2}{c}{\textbf{SpT}} & \textbf{Mass} & \textbf{Ref.} \\
 & Prim & Comp & & \\
\hline
06575703+6219197 & M4 & M5 & $\cdots$ & 1 \\
11240434+3808108 & M4.5 & M9.5 & 81 $\pm$ 5 M$_{Jup}$ & 2, 3 \\
15553178+3512028 & M4 & M7 & $\cdots$ & 4 \\
15594729+4403595 & M1.5 & M7.5$\pm$0.5 & 43 $\pm$ 9 M$_{Jup}$ & 5, 3 \\
19074283+3232396 & $\cdots$ &  \new{$\cdots$} & 0.42 $\pm$ 0.03 M$_{\odot}$\new{*} & 6\\
20231789+6710096 & M5 & M5 & $\cdots$ & 7 \\
\hline
\end{tabular}
\tablenotetext{}{\new{* This is the total system mass (M$_{tot}$)}}
\tablenotetext{}{[1] \cite{Newton14}, [2] \cite{Close03}, [3] \cite{Bowler15b}, [4] WDS catalog, [5] \cite{Janson12}, [6] \cite{Mann19}, [7] \cite{Law08}}
\end{table}

Below we summarize companions not characterized in the literature:

2MASS J06584690+2843004 is an M-type star at a \textit{Gaia} determined distance of 41.75 $\pm$ 1.2 pc. We detect a companion candidate at a separation of 0.88$\arcsec$. We found no match in \textit{Gaia} DR2 for the companion, so follow-up is needed to confirm whether it is physically associated. We do not detect it in the visible image so we place a minimum \textit{i'}--\textit{H} color limit of 2.07 mag. Using isochrone models, we estimate its mass to be 17--86 M$_{Jup}$ for an age range of 10--200 Myrs.

2MASS J12082885+1327090 is at a distance of 37.49 $\pm$ 0.11 pc. We find a physically associated companion at 1.8$\arcsec$, or a projected separation of 68.43 AU. Its \textit{i'--J} color is 3.04 mag corresponding to an estimated SpT of M6.6 from \cite{Kraus07} SEDs. By using PanSTARRS photometry we determine this object is red  with \textit{g--r} and \textit{z--y} colors of 2.07 and 0.96, respectively. Using the color and spectral type estimates from \cite{Best18}, this object would be a late M/early L-type object. Its isochrone mass range for 10-200 Myrs is 21-135 M$_{Jup}$.

2MASS J14192958+0254365 is an M5 star at a distance of 20.83 $\pm$ 0.04 pc. We detect a companion candidate at 0.5$\arcsec$ with a SNR of 6.7, just above our detection threshold of 5. We do not detect it in the visible image, thus placing a \textit{i'--J} color limit of $>$ 3.6 mag and late M spectral type. We do not detect the companion in \textit{Gaia} DR2.

2MASS J15471513+0149218 is at a distance of 18.39 $\pm$ 0.14 pc. We detect a companion candidate at a separation of 0.23$\arcsec$, with no detection in the visible camera, corresponding to an \textit{i'--J} color-limit of 5.2 mag. Using isochrone models, we estimate its mass to be 14--56 M$_{Jup}$ for an age range of 10--200 Myrs. PanSTARRS detects a very blue object at a projected separation of $\sim4\arcsec$ and PA of $\sim300^{\circ}$. It is unassociated according to \textit{Gaia} DR2 measurements and catalogued as a white dwarf by \cite{Bai18}. This blue object is too faint for both the visible and infrared Robo-AO cameras. 

2MASS J16304072-2018186 is a member of Sco-Cen and thus very young (5--15 Myrs) compared to most of our other targets \new{and at a distance of 182.82 $\pm$ 4.27 pc}. We detected two nearby objects, also detected by \textit{Gaia} DR2. The \textit{Gaia} data indicates that only one is a physically associated companion \new{at a projected separation of 551.63 AU} while the other one is a background object. These objects are visible in PanSTARRS data, which measures the physical companion as a very red object. \new{We obtained near-infrared spectra of both the primary star (2M1630-2018A) and the physically associated companion (2M1630-2018B) using the NASA Infrared Telescope Facility (IRTF) in prism mode on 2020 August 23 UT. We took 8 exposures for both objects in a ABBA pattern with 10~seconds and 120~seconds each for the primary and companion, respectively, and we contemporaneously observed a nearby A0V standard star HD~152071 for telluric correction. We reduced the data using \texttt{Spextool} version 4.1 \citep{Cushing04} and our resulting spectra have a median SNR of 215 per pixel for the primary and 55 per pixel for the companion in $J$ band. Comparing these objects' spectra with M-type spectral standards from \cite{Kirkpatrick10} in each of the $J$, $H$, and $K$ bands, we derive visual near-infrared spectral types of M3.5$\pm$1 and M5$\pm$1 for the primary and companion, respectively. Quantitative spectral types are not available for these objects as their H$_{2}$O-band spectral indices exceed the applicable range of \cite{AllersLiu13} and \cite{Zhang18} methods.}

\begin{figure*}[ht!]
\centering
\subfloat{\includegraphics[width=420pt]{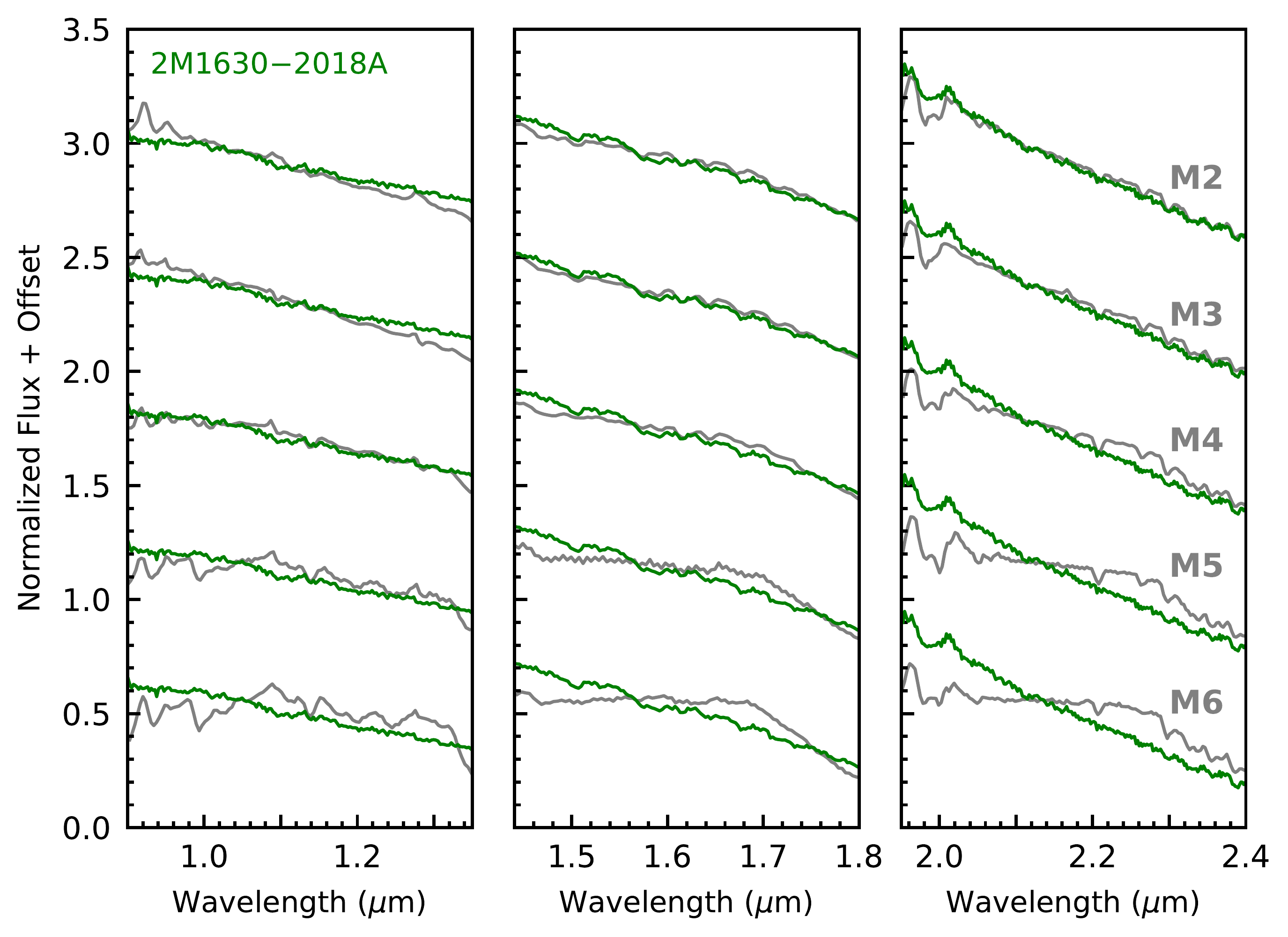}}\\
\subfloat{\includegraphics[width=420pt]{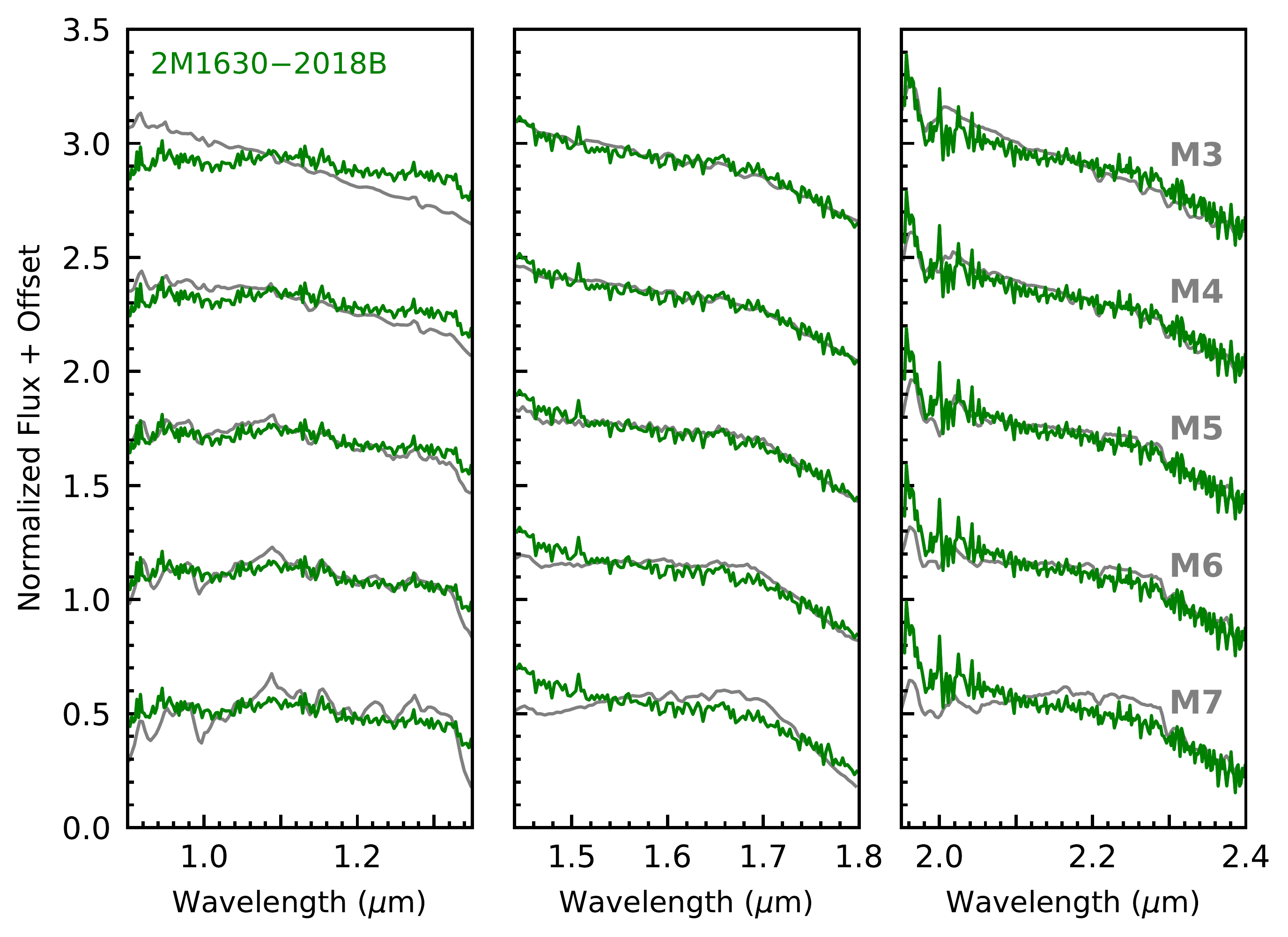}}
\caption{\new{IRTF/SpeX spectra of 2M1630A (top) and 2M1630B (bottom) as compared to the \cite{Kirkpatrick10} M-dwarf spectral standards (grey) in $J$, $H$, and $K$ bands. All these standards are normalized by the averaged flux of our targets within each band and their names are Gl 91 (M2), Gl 752A (M3), Gl 213 (M4), Gl 51 (M5),  LHS 1375 (M6), vB 8 (M7). We derive a visual near-infrared spectral type of M3.5$\pm$1 and M5$\pm$1 for the primary and the companion, respectively.}
    \label{fig:SpeX_2M1630A}}
\end{figure*}

\subsection{Survey Yields and Expectations}
\label{sec:yields}

\new{This is an ongoing survey and a detailed statistical analysis of these results is beyond the scope of this paper.} We calculate the \new{raw} multiplicity \new{detection frequency} of the \Nstars stars we observed, excluding detections ruled out as background objects, and including companions \new{where physical association is} to-be-determined. The fraction of observed stars with companion(s) is \new{24.1 $\pm$ 2.4\%} (half still need confirmation). \new{The breakdown of multiplicity fractions between the young field stars and Sco-Cen association stars are summarized in Table \ref{tab:detection_freqs}. Uncertainties are calculated following the binomial distribution for samples $<100$ stars and the Poisson distribution for the larger samples.} 

\begin{table}[h]
\centering
\caption{\new{Companion Detection Frequencies}
    \label{tab:detection_freqs}}
\begin{tabular}{lcrl}
\new{\textbf{Sample}} & \new{\textbf{\# stars}} & \multicolumn{2}{l}{\new{\textbf{\# stars with}}} \\
 & \new{\textbf{observed}} & \multicolumn{2}{l}{\new{\textbf{companion(s)}}} \\
\hline
\new{Total} & \new{\Nstars} & \new{103} & \new{(24.1 $\pm$ 2.4\%)} \\
\hline
\new{Total young field} & \new{403} & \new{99} & \new{(24.6 $\pm$ 2.5\%)} \\
\new{\quad LASSO young field} & \new{\Nstarsfield} & \new{70} & \new{(21.8 $\pm$ 2.6\%)} \\
\new{\quad Pre-LASSO young field} & \new{\NstarspreLASSO} & \new{29} & \new{(35$^{+6}_{-5}$\%)} \\
\new{Sco-Cen association} & \new{\NstarsScoCen} & \new{4} & \new{(17$^{+10}_{-5}$\%)} \\
\hline
\end{tabular}
\end{table}

Cool dwarfs have lower multiplicity fractions than higher mass stars \citep{Dupuy13} with fractions of $26\pm3$\% for stars $\new{0.1-}0.7$~M$_{\odot}$ \citep{Duchene13}, consistent with our results. \new{Our Sco-Cen sample is an order of magnitude smaller than our young field sample, so a detailed statistical comparison of our young field sample and our younger Sco-Cen sample is beyond the scope of this paper. We also note the difference in detection frequencies between our LASSO and pre-LASSO samples. The pre-LASSO sample was constructed using photometric distances, which would increase the chances of inadvertently including more distant binaries, thought to be less distant single stars. The fraction of higher order systems is 0.9 $\pm$ 0.5\%. \cite{Winters19} estimate a higher-order multiplicity rate of $\sim$5\%, from a companion search at separations of 2$\arcsec$ to 300$\arcsec$, which is over an order of magnitude larger than our field of view, where many of the wide-orbit tertiary companions would be.}

One bona fide substellar object and one companion at the substellar-stellar mass boundary were detected in our survey (2MASS J15594729+4403595 and 2MASS J11240434+3808108) and another $\sim$5 are potentially substellar. This gives us a detection rate of $0.5 - 1.5\%$. This is somewhat lower than the current overall frequency estimates of $1-4\%$ \citep{BowlerNielson18} for 13--75 M$_{Jup}$ companions. This is likely because we did not have the sensitivity necessary to detect substellar objects in all of our observations \new{(see Figure \ref{fig:detection_sensitivity})}. The dome-seeing fluctuated throughout the survey, affecting the AO performance. In addition, we do not have precise age estimates for most of our targets, especially those not part of young moving groups. \cite{Kastner18} discuss the UV-excess selection method for youth that we followed may not only be selecting stars as young as we expect. They investigated a sample of 400 low-mass (K and early-M type) stars expected to be young from UV-excess and with isochrone ages $\lesssim$80 Myr. However, a portion of those stars were fainter in the X-ray than expected, which could suggest they are not as young as expected. 

\section{Conclusion}
\label{sec:Conclusion}

The ongoing LASSO survey is one of the largest direct imaging surveys searching for wide-orbit substellar companions. The goal is to study the demographics of wide-orbit substellar companion populations in order to better understand their formation and evolution mechanisms. 

We have observed \Nstars  young, nearby, low-mass stars with Robo-AO on the 2.1-m telescope on Kitt Peak, Arizona, and on the UH 2.2-m telescope on Maunakea, Hawai`i, simultaneously in the visible and near-infrared. Our main findings are summarized below:

\begin{itemize}

\item We detected \Ncomps companion candidates and determined that \Ncompsbound are likely physically associated from \textit{Gaia} DR2 \new{and EDR3} common parallax and proper motion measurements \new{and a literature search}. Another \Ncompsunknown have yet to be confirmed, though we expect most of them to also be physically associated. The remaining \Ncompsbackground are background objects. 

\item \new{We were sensitive to substellar companions for 50\% of our LASSO field observations for separation ranges of 50--450 AU and for 90\% of our Sco-Cen observations for separation ranges of 210--570 AU.}

\item \new{Four} triple system candidates were detected, three of which have been previously reported in the literature.

\item We detected one bona fide brown dwarf, one companion at the threshold between brown dwarf and stellar mass, and another 5 companion candidates that will require follow-up observations to determine their nature.

\item The range of projected separations spans 2--1101 AU and masses $\gtrsim$43 M$_{Jup}$. 

\item We investigated accelerations calculated from \textit{Hipparcos-Gaia} proper motion measurements and found that a higher fraction (58\new{$^{+12}_{-14}$}\%) of stars with companions are accelerating compared to stars without detected companions (23\new{$^{+11}_{-6}$}\%). These accelerating stars with detected companions will allow us to calculate dynamical masses with future orbit monitoring.

\item \new{Our  multiplicity fractions are 24.1~$\pm$~2.4\% for the entire sample, 24.6 $\pm$ 2.5 \% for the young field stars, and 17$^{+10}_{-5}$\% for the Sco-Cen sample.}

\end{itemize}

\acknowledgments

The authors are honored to be permitted to conduct astronomical research on Iolkam Du'ag (Kitt Peak), a mountain with particular significance to the Tohono O'odham Nation. The authors also wish to recognize and acknowledge the very significant cultural role and reverence that the summit of Maunakea has always had within the indigenous Hawaiian community. We are most fortunate to have the opportunity to conduct observations from both mountains.

We are grateful to the Kitt Peak National Observatory and UH88" staff for their support of Robo-AO on the 2.1-m and 2.2-m telescopes, respectively. We thank Shri Kulkarni for his sustained backing of Robo-AO through all its iterations\new{,} Dani Atkinson for help understanding the intricacies of SAPHIRA detectors\new{, and} Bo Reipurth for valuable discussions and comments on the manuscript. We are grateful to Adwin Boogert for observing 2MASS J16304072-2018186 and its companion \new{with} IRTF/SpeX.

The Robo-AO system is supported by collaborating partner institutions, the California Institute of Technology and the Inter-University Centre for Astronomy and Astrophysics, and by the National Science Foundation under Grant Nos. AST-0906060, AST-0960343, and AST-1207891, by the Mount Cuba Astronomical Foundation, and by a gift from Samuel Oschin. As part of the development of Robo-AO-2, Robo-AO at the UH~2.2-m telescope system is supported by the National Science Foundation under Grant No. AST-1712014, the State of Hawaii Capital Improvement Projects, and by a gift from the Lumb Family. 

\new{M.C.L. acknowledges support from National Science Foundation grant AST-1518339.}

B.P.B. acknowledges support from the National Science Foundation grant AST-1909209.

This work has made use of data from the European Space Agency (ESA) mission {\it Gaia} (\url{https://www.cosmos.esa.int/gaia}), processed by the {\it Gaia} Data Processing and Analysis Consortium (DPAC, \url{https://www.cosmos.esa.int/web/gaia/dpac/consortium}). Funding for the DPAC has been provided by national institutions, in particular the institutions participating in the {\it Gaia} Multilateral Agreement.

This research has made use of the VizieR catalogue access tool, CDS, Strasbourg, France (DOI: 10.26093/cds/vizier). The original description of the VizieR service was published in A\&AS 143, 23.

This research has made use of the SIMBAD database, operated by Centre des Donn\'ees Stellaires (Strasbourg, France), and bibliographic references from the Astrophysics Data System maintained by SAO/NASA. 

This publication has made use of data products from the Two Micron All Sky Survey, which is a joint project of the University of Massachusetts and the Infrared Processing and Analysis Center/California Institute of Technology, funded by the National Aeronautics and Space Administration and the National Science Foundation.

This research has made use of the SVO Filter Profile Service (\url{http://svo2.cab.inta-csic.es/theory/fps/}) supported from the Spanish MINECO through grant AYA2017-84089

This research has made use of the Washington Double Star Catalog maintained at the U.S. Naval Observatory.

\facility{KPNO:2.1m (Robo-AO), UH:2.2m (Robo-AO)\new{, IRTF (SpeX)}}
\software{\new{ExoDMC (v1.1b; \citealt{Bonavita2020}), Spextool (v4.1; \citealt{Cushing04})}}

\bibliography{references.bib}{}
\bibliographystyle{aasjournal}

\appendix

Table \ref{tab:comps_measurements} lists Robo-AO companion detection measurements and, when available, \textit{Gaia} DR2 measurements. Tables \ref{tab:comp_properties_binaries_bound}, \ref{tab:comp_properties_binaries_unknown}, and \ref{tab:comp_properties_triples} list the properties calculated for physically associated binaries, unconfirmed binary candidates, and triple systems, respectively. Table \ref{tb:catalog_matches} lists the Robo-AO detected companions that are found in other catalogs (\citealt{Lamman20,CortesContreras17,Bowler19}, and WDS). Finally, Table \ref{tab:obs_all} lists all observed targets reported in this work.

\begin{longrotatetable}
\begin{deluxetable*}{lcccccccc|cccccc}
\tablecaption{Companion Candidate Detection Measurements}
\tablewidth{700pt}
\tabletypesize{\scriptsize}
\tablehead{
\multicolumn{9}{c|}{\textbf{Robo-AO}} & \multicolumn{6}{c}{\textbf{\textit{Gaia} DR2}} \\ 
\colhead{Primary 2MASS ID} & \colhead{Separation} & \colhead{P. A.} & \multicolumn{3}{c}{Magnitude Difference ($\Delta$m)} & \multicolumn{3}{c|}{Detection SNR} & & & & \colhead{Phys.} & RUWE & \colhead{HIP-\textit{Gaia}} \\ 
& \colhead{($\arcsec$)} & \colhead{($^{\circ}$)} & \colhead{\textit{i'}} & \colhead{\textit{J}} & \colhead{\textit{H}} & \textit{i'} & \textit{J} & \colhead{\textit{H}} \vline & \colhead{$\Delta$m$_{\textit{G}}$} & \colhead{$\Delta \pi/\pi$} & \colhead{$\Delta \mu/\mu$} & \colhead{assoc.} & \colhead{prim. (comp)} & \colhead{prim. $\chi^2_{acc}$} }
\startdata
00024011+3821453 B & 1.41 $\pm$ 0.03 & 26.5 $\pm$ 0.6 & 0.2 $\pm$ 0.04 & $\cdots$ & 0.14 $\pm$ 0.1 & 390 & $\cdots$ & 119 & 0.17 & 10$^{-2}$ & 10$^{-2}$ & yes & 4.1 (10.7) & $\cdots$\\  
00175864+2057192 B & 9.33 $\pm$ 0.02 & 232.4 $\pm$ 0.1 & 0.55 $\pm$ 0.0 & 0.3 $\pm$ 0.13 & $\cdots$ & 653 & 195 & $\cdots$ & 0.59 & 10$^{-3}$ & 10$^{-3}$ & yes & 1.0 (1.2) & $\cdots$\\  
00213183+1934253 B & 4.86 $\pm$ 0.02 & 224.3 $\pm$ 0.4 & 3.1 $\pm$ 0.02 & $\cdots$ & 2.25 $\pm$ 0.1 & 483 & $\cdots$ & 207 & 3.1 & 10$^{-3}$ & 0.1 & yes & 1.3 (1.2) & $\cdots$\\  
00233382+0947356 B & 3.61 $\pm$ 0.01 & 316.4 $\pm$ 0.2 & $\cdots$ & 0.09 $\pm$ 0.05 & $\cdots$ & $\cdots$ & 116 & $\cdots$ & -0.29 & 10$^{-3}$ & 10$^{-2}$ & yes & 1.2 (1.3) & $\cdots$\\  
00360847+4530575 B & 0.52 $\pm$ 0.02 & 120.9 $\pm$ 1.3 & $\cdots$ & 1.65 $\pm$ 0.08 & $\cdots$ & $\cdots$ & 19 & $\cdots$ & $\cdots$ & $\cdots$ & $\cdots$ & $\cdots$ & $\cdots$ & $\cdots$\\  
01563544+1228047 B & 2.89 $\pm$ 0.02 & 240.4 $\pm$ 0.4 & 1.24 $\pm$ 0.01 & 0.96 $\pm$ 0.08 & $\cdots$ & 344 & 116 & $\cdots$ & 1.46 & 0.2 & 10$^{-3}$ & yes & 5.5 (1.0) & $\cdots$\\  
01564996+4142303 B & 0.19 $\pm$ 0.03 & 315.9 $\pm$ 5.2 & 2.36 $\pm$ 0.2 & 2.23 $\pm$ 0.1 & $\cdots$ & 7 & 4 & $\cdots$ & $\cdots$ & $\cdots$ & $\cdots$ & $\cdots$ & 1.6 & $\cdots$\\  
01584363+3515281 B & 1.85 $\pm$ 0.01 & 86.3 $\pm$ 0.3 & $\cdots$ & 0.2 $\pm$ 0.07 & $\cdots$ & $\cdots$ & 131 & $\cdots$ & -0.06 & 10$^{-2}$ & 0.1 & yes & 2.4 (1.5) & $\cdots$\\
\enddata
\tablenotetext{}{(SC) Sco-Cen member}
\tablenotetext{a}{Although 2MASS 16304072-2018186 has a detection significance below our threshold, we have confirmed that it is real from PanSTARRS observations and a match in \textit{Gaia} DR2.}
\tablenotetext{b}{2MASS J02514973+2929131 \& 2MASS J13414631+5815197: \cite{CortesContreras17}; 2MASS J10364483+1521394: \cite{Calissendorff17}; 2MASS J19535443+4424541: \cite{Tokovinin17}; 2MASS J20301067+2650344: \cite{Malkov12}; 2MASS J19074283+3232396 \& 2MASS J23350028+0136193: \cite{Mann19}}
\tablenotetext{c}{From Gaia EDR3 astrometry}
\label{tab:comps_measurements}
\end{deluxetable*}
\end{longrotatetable}

\begin{longrotatetable}
\begin{deluxetable*}{lcccccccccc}
\tablecaption{Binary Companion Properties: Physically Associated}
\tablewidth{700pt}
\tabletypesize{\scriptsize}
\tablehead{
\colhead{2MASS ID} & \colhead{Separation} & \multicolumn{3}{c}{Apparent Magnitude} & \multicolumn{3}{c}{Absolute Magnitude} & \multicolumn{2}{c}{vis-IR color} & \colhead{Mass$^a$} \\ 
& \colhead{(AU)} & \colhead{\textit{i'}} & \colhead{\textit{J}} & \colhead{\textit{H}} & \textit{i'} & \textit{J} & \colhead{\textit{H}} & \textit{i'-J} & \textit{i'-H}  & \colhead{(M$_{Jup}$)} } 
\startdata 
00024011+3821453 B & 58.52 $\pm$ 1.45 & 11.99 $\pm$ 0.04 & $\cdots$ & 9.34 $\pm$ 0.1 & 8.91 $\pm$ 0.04 & $\cdots$ & 6.26 $\pm$ 0.35 & $\cdots$ & 2.65 $\pm$ 0.11 & 175 -- 495\\  
00175864+2057192 B & 259.33 $\pm$ 0.74 & 10.05 $\pm$ 0.0 & 8.99 $\pm$ 0.13 & $\cdots$ & 7.83 $\pm$ 0.0 & 6.77 $\pm$ 0.13 & $\cdots$ & 1.06 $\pm$ 0.13 & $\cdots$ & 172 -- 504\\  
00213183+1934253 B & 233.75 $\pm$ 1.5 & 14.18 $\pm$ 0.02 & $\cdots$ & 11.18 $\pm$ 0.1 & 10.77 $\pm$ 0.02 & $\cdots$ & 7.77 $\pm$ 0.14 & $\cdots$ & 3.0 $\pm$ 0.1 & 64 -- 247\\  
00233382+0947356 B & 141.53 $\pm$ 0.55 & $\cdots$ & 9.89 $\pm$ 0.05 & $\cdots$ & $\cdots$ & 6.92 $\pm$ 0.07 & $\cdots$ & $>$ 4.53 & $\cdots$ & 154 -- 477\\  
01563544+1228047 B & 133.42 $\pm$ 1.1 & 11.18 $\pm$ 0.01 & 9.26 $\pm$ 0.08 & $\cdots$ & 7.86 $\pm$ 0.01 & 5.93 $\pm$ 0.09 & $\cdots$ & 1.92 $\pm$ 0.08 & $\cdots$ & $>$ 312\\  
01584363+3515281 B & 83.16 $\pm$ 0.69 & $\cdots$ & 9.85 $\pm$ 0.07 & $\cdots$ & $\cdots$ & 6.58 $\pm$ 0.11 & $\cdots$ & $>$ 4.88 & $\cdots$ & 195 -- 535\\  
02022823+1034533 B & 13.46 $\pm$ 0.31 & 11.12 $\pm$ 0.07 & $\cdots$ & 8.21 $\pm$ 0.15 & 10.31 $\pm$ 0.07 & $\cdots$ & 7.39 $\pm$ 0.26 & $\cdots$ & 2.91 $\pm$ 0.16 & 83 -- 303\\  
02170213+2719305 B & 120.22 $\pm$ 0.88 & 10.38 $\pm$ 0.01 & 9.95 $\pm$ 0.07 & $\cdots$ & 6.54 $\pm$ 0.01 & 6.11 $\pm$ 0.09 & $\cdots$ & 0.43 $\pm$ 0.07 & $\cdots$ & 275 -- 614\\
\enddata 
\tablenotetext{}{(SC) Sco-Cen member} 
\tablenotetext{a}{Estimated mass ranges are from evolutionary isochrone models \citep{Chabrier00,Baraffe15} calculated for ages 10-200~Myr, except Sco-Cen members were calculated for 5-15~Myr.} 
\label{tab:comp_properties_binaries_bound} 
\end{deluxetable*}\end{longrotatetable}
\begin{longrotatetable}
\begin{deluxetable*}{lccccccccccc}
\tablecaption{Binary Companion Candidate Properties: Physical Association To Be Confirmed}
\tablewidth{700pt}
\tabletypesize{\scriptsize}
\tablehead{
\colhead{2MASS ID} & \colhead{Separation} & \multicolumn{3}{c}{Apparent Magnitude} & \multicolumn{3}{c}{Absolute Magnitude} & \multicolumn{2}{c}{vis-IR color} & \colhead{Mass$^a$} & \colhead{P($N_{background} \geq 1$)} \\ 
& \colhead{(AU)} & \colhead{\textit{i'}} & \colhead{\textit{J}} & \colhead{\textit{H}} & \textit{i'} & \textit{J} & \colhead{\textit{H}} & \textit{i'-J} & \textit{i'-H}  & \colhead{(M$_{Jup}$)}  & \colhead{(\%)} } 
\startdata 
00360847+4530575 B & 11.03 $\pm$ 0.37 & $\cdots$ & 9.82 $\pm$ 0.08 & $\cdots$ & $\cdots$ & 8.18 $\pm$ 0.08 & $\cdots$ & $\cdots$ & $\cdots$ & 64 -- 259 & 4.88 $\pm$ 0.11\\  
01564996+4142303 B & 5.92 $\pm$ 0.9 & 13.25 $\pm$ 0.2 & 11.94 $\pm$ 0.1 & $\cdots$ & 10.82 $\pm$ 0.2 & 9.52 $\pm$ 0.12 & $\cdots$ & 1.3 $\pm$ 0.22 & $\cdots$ & 20 -- 123 & 3.97 $\pm$ 0.1\\  
02154870+2458501 B & 33.64 $\pm$ 0.84 & 12.56 $\pm$ 0.03 & $\cdots$ & 9.61 $\pm$ 0.19 & 9.12 $\pm$ 0.03 & $\cdots$ & 6.16 $\pm$ 0.23 & $\cdots$ & 2.95 $\pm$ 0.2 & 186 -- 510 & 1.13 $\pm$ 0.04\\  
02284694+1538535 B & 25.76 $\pm$ 0.86 & 12.23 $\pm$ 0.03 & $\cdots$ & 9.14 $\pm$ 0.14 & 9.51 $\pm$ 0.03 & $\cdots$ & 6.42 $\pm$ 0.16 & $\cdots$ & 3.09 $\pm$ 0.14 & 157 -- 469 & 0.96 $\pm$ 0.03\\  
02405251+4452365 B & 8.72 $\pm$ 0.38 & 11.68 $\pm$ 0.07 & 10.89 $\pm$ 0.51 & $\cdots$ & 9.95 $\pm$ 0.07 & 9.16 $\pm$ 0.55 & $\cdots$ & 0.79 $\pm$ 0.51 & $\cdots$ & 23 -- 150 & 5.41 $\pm$ 0.11\\  
02453008+2620233 B & 11.55 $\pm$ 1.0 & 12.84 $\pm$ 0.23 & 11.04 $\pm$ 0.15 & $\cdots$ & 8.78 $\pm$ 0.23 & 6.99 $\pm$ 1.28 & $\cdots$ & 1.79 $\pm$ 0.28 & $\cdots$ & 147 -- 466 & 2.01 $\pm$ 0.07\\  
03051963+2131219 B & 22.81 $\pm$ 0.77 & 13.42 $\pm$ 0.08 & 10.71 $\pm$ 0.31 & $\cdots$ & 10.73 $\pm$ 0.08 & 8.02 $\pm$ 0.43 & $\cdots$ & 2.71 $\pm$ 0.32 & $\cdots$ & 73 -- 284 & 1.74 $\pm$ 0.06\\  
03051963+2131219 B & 23.03 $\pm$ 0.6 & 12.67 $\pm$ 0.1 & $\cdots$ & 9.49 $\pm$ 0.15 & 9.97 $\pm$ 0.1 & $\cdots$ & 6.8 $\pm$ 0.34 & $\cdots$ & 3.17 $\pm$ 0.18 & 122 -- 407 & 1.24 $\pm$ 0.04\\
\enddata 
\tablenotetext{}{(SC) Sco-Cen member} 
\tablenotetext{a}{Estimated mass ranges are from evolutionary isochrone models \citep{Chabrier00,Baraffe15} calculated for ages 10-200~Myr, except Sco-Cen members were calculated for 5-15~Myr.} 
\label{tab:comp_properties_binaries_unknown} 
\end{deluxetable*}\end{longrotatetable}
\begin{longrotatetable}
\begin{deluxetable*}{lcccccccccccc}
\tablecaption{Triple Companion Candidate Properties}
\tablewidth{700pt}
\tabletypesize{\scriptsize}
\tablehead{
\colhead{2MASS ID} & \colhead{Separation} & \multicolumn{3}{c}{Apparent Magnitude} & \multicolumn{3}{c}{Absolute Magnitude} & \multicolumn{2}{c}{vis-IR color} & \colhead{Mass$^a$} & \colhead{Phys.} & \colhead{P($N_{background} \geq 1$)} \\ 
& \colhead{(AU)} & \colhead{\textit{i'}} & \colhead{\textit{J}} & \colhead{\textit{H}} & \textit{i'} & \textit{J} & \colhead{\textit{H}} & \textit{i'-J} & \textit{i'-H} & \colhead{(M$_{Jup}$)} & \colhead{assoc.} & \colhead{(\%)}}\startdata 
05242572+1922070 B & 113.22 $\pm$ 1.32 & 13.49 $\pm$ 0.37 & 11.67 $\pm$ 0.09 & $\cdots$ & 9.62 $\pm$ 0.37 & 7.81 $\pm$ 0.21 & $\cdots$ & 1.82 $\pm$ 0.38 & $\cdots$ & 84 -- 319 & yes & \multirow{2}{*}{7.95 $\pm$ 0.13}\\  
05242572+1922070 C & 20.92 $\pm$ 0.75 & $\cdots$ & 11.34 $\pm$ 0.29 & $\cdots$ & $\cdots$ & 7.51 $\pm$ 0.52 & $\cdots$ & $>$ 3.58 & $\cdots$ & 102 -- 375 & $\cdots$ & \\  
10364483+1521394 B & 17.22 $\pm$ 0.31 & 11.98 $\pm$ 0.03 & 9.46 $\pm$ 0.08 & $\cdots$ & 10.5 $\pm$ 0.03 & 7.98 $\pm$ 0.09 & $\cdots$ & 2.52 $\pm$ 0.08 & $\cdots$ & 74 -- 290 & yes & \multirow{2}{*}{$\cdots$}\\  
10364483+1521394 C & 14.47 $\pm$ 0.28 & 12.05 $\pm$ 0.01 & 9.43 $\pm$ 0.08 & $\cdots$ & 10.57 $\pm$ 0.01 & 7.95 $\pm$ 0.09 & $\cdots$ & 2.62 $\pm$ 0.08 & $\cdots$ & 76 -- 295 & yes & \\  
19535443+4424541 B & 1.93 $\pm$ 0.05 & 12.03 $\pm$ 0.03 & 9.56 $\pm$ 0.1 & $\cdots$ & 13.67 $\pm$ 0.03 & 11.2 $\pm$ 0.1 & $\cdots$ & 2.47 $\pm$ 0.11 & $\cdots$ & 13 -- 54 & yes & \multirow{2}{*}{$\cdots$}\\  
19535443+4424541 C & 27.38 $\pm$ 0.11 & 9.35 $\pm$ 0.02 & 8.16 $\pm$ 0.07 & $\cdots$ & 11.01 $\pm$ 0.02 & 9.82 $\pm$ 0.07 & $\cdots$ & 1.19 $\pm$ 0.07 & $\cdots$ & 18 -- 103 & yes & \\  
23350028+0136193 B & 17.69 $\pm$ 0.26 & $\cdots$ & 8.38 $\pm$ 0.17 & $\cdots$ & $\cdots$ & 6.78 $\pm$ 0.17 & $\cdots$ & $>$ 3.48 & $\cdots$ & 170 -- 501 & yes & \multirow{2}{*}{$\cdots$}\\  
23350028+0136193 C & 25.38 $\pm$ 0.28 & $\cdots$ & 8.1 $\pm$ 0.18 & $\cdots$ & $\cdots$ & 6.51 $\pm$ 0.18 & $\cdots$ & $>$ 3.8 & $\cdots$ & 205 -- 547 & yes & \\
\enddata 
\tablenotetext{}{(SC) Sco-Cen member} 
\tablenotetext{a}{Estimated mass ranges are from evolutionary isochrone models \citep{Chabrier00,Baraffe15} calculated for ages 10-200~Myr, except Sco-Cen members were calculated for 5-15~Myr.} 
\label{tab:comp_properties_triples} 
\end{deluxetable*}\end{longrotatetable}

\startlongtable
\begin{deluxetable*}{c|cccccc}
\tablecaption{Companion Candidates Found in Catalogs}
\tablehead{
\colhead{Primary 2MASS ID} \vline & \colhead{Catalog} & \colhead{Obs Date} & \colhead{Separation} & \colhead{P. A.}  & \colhead{Mass} & \colhead{SpT}\\ 
& & \colhead{(UT)} & \colhead{($\arcsec$)} & \colhead{($^{\circ}$)} & \colhead{M$_{\odot}$} & }
\startdata
01584363+3515281 B & This Work & 2017 Oct 29 & 1.85 $\pm$ 0.01 & 86.3 $\pm$ 0.3 & $\cdots$ & $\cdots$ \\  
 & WDS  & 2015 & 1.9 & 271 & $\cdots$ & $\cdots$\\  
02022823+1034533$^a$ C & This Work & 2019 Oct 16 & 3.99 $\pm$ 0.02 & 185.9 $\pm$ 0.3 & $\cdots$ & $\cdots$ \\  
 & WDS  & 2014 & 4.7 & 191 & $\cdots$ & $\cdots$\\  
02170213+2719305 B & This Work & 2017 Oct 28 & 2.05 $\pm$ 0.01 & 183.6 $\pm$ 0.4 & $\cdots$ & $\cdots$ \\  
 & WDS  & 2015 & 2.1 & 183 & $\cdots$ & $\cdots$\\  
02284694+1538535 B & This Work & 2019 Oct 16 & 0.73 $\pm$ 0.02 & 137.9 $\pm$ 1.2 & $\cdots$ & $\cdots$ \\  
 & B19 & 2014 Nov 07 & 0.84 $\pm$ 0.04 & 147 $\pm$ 3 & $\cdots$ & $\cdots$ \\
\enddata
\tablenotetext{}{(SC) Sco-Cen member}
\tablenotetext{}{\textbf{Note:} L20: \cite{Lamman20}; C17: \cite{CortesContreras17}; B19: \cite{Bowler19}; WDS: Washington Double Star catalog}
\tablenotetext{a}{We determined this companion candidate to be a background object from \textit{Gaia} DR2 measurements (See Table \ref{tab:comps_measurements}).}
\tablenotetext{b}{The WDS lists the separation and PA of the BC components of the system while we list the separation and PA for AB and AC.}
\label{tb:catalog_matches}
\end{deluxetable*}

\begin{longrotatetable}
\begin{deluxetable*}{lcccccccccc}
\tablecaption{All Observed Targets}
\tablehead{
\colhead{2MASS ID} & \colhead{RA} & \colhead{Dec} & \colhead{m\textit{$_J$}}  & \colhead{Obs. Date} & \colhead{Seeing} & \colhead{Reduction} & \colhead{Filter(s)} & \colhead{IR Contrast Limits ($\Delta$m)} & \colhead{Max Sep.} & \colhead{Comp.}\\ 
& \colhead{J2000} & \colhead{J2000} & \colhead{(mag)} & \colhead{(UT)} & \colhead{($\arcsec$)} & \colhead{Pipeline} & & \colhead{[0.5$\arcsec$, 1.0$\arcsec$, 2.0$\arcsec$, 3.0$\arcsec$]} & \colhead{(AU)} & \colhead{Candidate?} }
\startdata
00004701+1624101 & 00:00:47.01 & +16:24:10.1 & 9.32 & 2017 Nov 06 & 2.01 & All frames & \textit{i', J} & 1.4, 3.8, 4.9, 5.8 & 391 & $\cdots$\\  
00024011+3821453 & 00:02:40.12 & +38:21:45.31 & 9.71 & 2019 Oct 12 & 1.24 & All frames & \textit{i', H} & 3.5, 4.8, 5.0, 7.9 & 244 & Y\\  
00082730+1725274 & 00:08:27.3 & +17:25:27.49 & 7.81 & 2019 Oct 13 & $\cdots$ & All frames & \textit{i', H} & 2.7, 3.9, 6.2, 8.0 & 128 & $\cdots$\\  
00175864+2057192 & 00:17:58.64 & +20:57:19.2 & 8.69 & 2017 Oct 30 & 1.38 & All frames & \textit{i', J} & 1.5, 4.4, 6.4, 7.7 & 228 & Y\\  
00205719+1738160 & 00:20:57.19 & +17:38:16.08 & 8.37 & 2019 Oct 16 & 1.01 & 90\% GenSTAC & \textit{i', H} & 2.8, 3.0, 4.4, 5.9 & 176 & $\cdots$\\  
00213183+1934253 & 00:21:31.84 & +19:34:25.31 & 9.53 & 2019 Oct 16 & 0.97 & All frames & \textit{i', H} & 3.0, 5.4, 7.0, 8.7 & 283 & Y\\  
00233382+0947356 & 00:23:33.82 & +09:47:35.6 & 9.79 & 2017 Nov 06 & 1.53 & All frames & \textit{i', J} & 1.8, 3.4, 5.0, 4.4 & 321 & Y\\  
00360847+4530575 & 00:36:08.47 & +45:30:57.5 & 8.17 & 2017 Oct 27 & $\cdots$ & All frames & \textit{J} & 0.2, 2.5, 4.2, 5.2 & 174 & Y\\
\enddata
\tablenotetext{}{(SC) Sco-Cen member}
\label{tab:obs_all}
\end{deluxetable*}
\end{longrotatetable}

\end{document}